\documentclass[aps,prd,twocolumn,10pt]{revtex4-1}
\usepackage{amssymb}
\usepackage{graphicx}
\usepackage{amsmath}
\usepackage{hyperref}
\usepackage{subfigure}
\usepackage{multirow}
\usepackage{setspace}
\usepackage{verbatim}
\usepackage{float}
\usepackage{color}
\usepackage{ulem}
\usepackage[utf8]{inputenc}

\begin{document}

\title{No-go guide for late-time solutions to the Hubble tension: Matter perturbations}

\author{Rong-Gen Cai$^{1,2,3}$}

\author{Zong-Kuan Guo$^{1,2,3}$}

\author{Shao-Jiang Wang$^{1}$}
\email{schwang@itp.ac.cn (corresponding author)}

\author{Wang-Wei Yu$^{1,3}$}
\email{yuwangwei@mail.itp.ac.cn (corresponding author)}

\author{Yong Zhou$^{1}$}

\affiliation{$^1$CAS Key Laboratory of Theoretical Physics, Institute of Theoretical Physics, Chinese Academy of Sciences, Beijing 100190, China}
\affiliation{$^2$School of Fundamental Physics and Mathematical Sciences, Hangzhou Institute for Advanced Study (HIAS), University of Chinese Academy of Sciences (UCAS), Hangzhou 310024, China}
\affiliation{$^3$School of Physical Sciences, University of Chinese Academy of Sciences (UCAS), Beijing 100049, China}

\begin{abstract}
The Hubble tension seems to be a crisis with $\sim5\sigma$ discrepancy between the most recent local distance ladder measurement from type Ia supernovae calibrated by Cepheids and the global fitting constraint from the cosmic microwave background data. To narrow down the possible late-time solutions to the Hubble tension, we have used in a recent study [Phys. Rev. D 105, L021301 (2022)] an improved inverse distance ladder method calibrated by the absolute measurements of the Hubble expansion rate at high redshifts from the cosmic chronometer data, and found no appealing evidence for new physics at the late time beyond the $\Lambda$CDM model characterized by a parametrization based on the cosmic age. In this paper, we further investigate the perspective of this improved inverse distance ladder method by including the late-time matter perturbation growth data. Independent of the dataset choices, model parametrizations, and diagnostic quantities ($S_8$ and $S_{12}$), the new physics at the late time beyond the $\Lambda$CDM model is strongly disfavored so that the previous late-time no-go guide for the Hubble tension is further strengthened.
\end{abstract}
\maketitle

\section{Introduction} 

The mismatch between the local and global values of the Hubble constant has been growing over the recent years from the Hubble discrepancy~\cite{Freedman:2010xv,Freedman:2017yms} to the Hubble tension~\cite{Bernal:2016gxb,Verde:2019ivm,Knox:2019rjx,Riess:2020sih,DiValentino:2020zio,DiValentino:2021izs}  and probably knocking the threshold of the Hubble crisis with the $\sim 5\sigma$ claim from the most recent measurement  $H_0=73.04\pm1.04$ km/s/Mpc~\cite{Riess:2021jrx} with unprecedented $\sim1$ km/s/Mpc uncertainty. Compared to the model-dependent constraint $H_0=67.27\pm0.60$ km/s/Mpc from Planck+$\Lambda$-cold-dark-matter ($\Lambda$CDM)~\cite{Planck:2018vyg}, the local distance ladder measurements from Hubble-flow type Ia supernovae (SNe Ia) calibrated by, for example, Cepheids~\cite{Riess:2016jrr,Riess:2018byc,Riess:2018uxu,Riess:2019cxk,Riess:2020fzl,Riess:2021jrx} are quasi-model-independent~\cite{Dhawan:2020xmp} provided that the absolute magnitude $M_B$ of SN Ia is measured \textit{a priori}. Changing the calibrators from Cepheids to the tip of the red giant branch (TRGB)~\cite{Freedman:2019jwv,Yuan:2019npk,Freedman:2020dne,Soltis:2020gpl,Freedman:2021ahq} might reduce the significance of the Hubble tension, however, combining Cepheid and TRGB consistently only slightly lowers the Hubble constant to $H_0=72.53\pm0.99$ km/s/Mpc~\cite{Riess:2021jrx}, therefore, the Hubble tension is still observationally pronounced to date.

The above local distance ladder (LDL) measurements from SNe+$M_B$ put strong constraints on the late-time models~\cite{Dhawan:2020xmp} with the inferred $H_0$ barely deviated from that of $\Lambda$CDM model, however, a noticeable exception comes from the late-time phantom transition model ~\cite{Mortonson:2009qq} with a rapid change in $H(z)$ at a lower redshift than the Hubble-flow SNe Ia. Although such a late-time phantom transition model could escape from the LDL constraint by raising the $H_0$ value without jeopardizing the Hubble-flow SNe Ia~\cite{Mortonson:2009qq}, no strong evidence for such a late-time phantom transition was found as shown in Ref.~\cite{Dhawan:2020xmp}. Furthermore, an inconsistency also shows up when applying the inverse distance ladder (IDL) method~\cite{Cuesta:2014asa,Heavens:2014rja,Aubourg:2014yra,Verde:2016ccp,Alam:2016hwk,Verde:2016wmz,Macaulay:2018fxi,Feeney:2018mkj,eBOSS:2020yzd} with a geometric calibration from baryon acoustic oscillations (BAO)~\cite{BOSS:2016wmc,eBOSS:2020yzd,DES:2021esc}. Note that the traditional IDL method from $r_d^\mathrm{Planck}$+BAO+SNe+$M_B$ requires additional $M_B$ prior and $r_d$ prior~\cite{Vonlanthen:2010cd,Audren:2012wb,Audren:2013nwa,Cuesta:2014asa,Verde:2016ccp,Bernal:2016gxb,Verde:2016wmz,Aylor:2018drw} from Planck+$\Lambda$CDM. Here, the $r_d$ prior is unharmful for constraining the late-time models since the sound horizon $r_d^\mathrm{Planck}\approx147$ Mpc only depends on the early-Universe evolution (hence insensitive to the late-time physics), however, the $M_B$ prior is crucial for the IDL method to discriminate the late-time models. The aforementioned inconsistency is two-folds as we elaborate below:

First, if the $M_B$ prior is fixed at the same value $M_B\approx -19.2$ mag for both LDL (SNe+$M_B$) and IDL  ($r_d^\mathrm{Planck}$+BAO+SNe+$M_B$), the inferred $H_0^\mathrm{LDL}\approx74$ km/s/Mpc and   $H_0^\mathrm{IDL}\approx68$ km/s/Mpc are in tension unless some early-Universe modification lowers the sound horizon down to $r_d^\mathrm{early}\approx137$ Mpc. This is usually the alternative view on the $H_0$ tension as the $r_d$ tension, and this is also the usual no-go argument ~\cite{Lemos:2018smw} for the general late-time solutions to the Hubble tension.  Second, for the special late-time solution like the phantom transition model, if leaving $M_B$ as a free parameter in both LDL (SNe) and IDL ($r_d^\mathrm{Planck}$+BAO+SNe), then not only the inferred distributions on $H_0^\mathrm{LDL}$ and $H_0^\mathrm{IDL}$ are in mild tension, but also the inferred distributions on $M_B^\mathrm{LDL}$ and $M_B^\mathrm{IDL}$ are in tension as well.  This is the recent intriguing view on the $H_0$ tension as the $M_B$ tension~\cite{Benevento:2020fev,Camarena:2021jlr,Efstathiou:2021ocp,Cai:2021weh}. In particular, if one reuses the inferred $M_B^\mathrm{IDL}$ distribution as a prior into the LDL by SNe+$M_B^\mathrm{IDL}$, then the inferred $H_0^\mathrm{LDL+IDL}$ distribution is also in larger tension~\cite{Efstathiou:2021ocp} with the $H_0^\mathrm{LDL}$ distribution than the pure $H_0^\mathrm{IDL}$ distribution. 
Therefore, the IDL method largely rules out the late-time solutions to the Hubble tension, at least for those homogeneous modifications (except for the inhomogeneous solutions from the interacting dark energy model~\cite{DiValentino:2019ffd} and chameleon dark energy model~\cite{Cai:2021wgv}). However, there are two major drawbacks in the traditional IDL method:

First, the input $r_d$ prior encodes the early-Universe evolution, although it is innocent for late-time model-selection, it would be more appealing to eliminate the use of a $r_d$ prior. For example, Ref.~\cite{Arendse:2019hev} proposed to use the strong lensing time delay (SLTD) from H0LiCOW measurement~\cite{Wong:2019kwg} on the time-delay distance $D_{\Delta t}$ to calibrate the IDL (namely SLTD+BAO+SNe+$M_B$). However, the SLTD measurements (for example, the TDCOSMO+SLACS sample~\cite{Birrer:2020tax}) are highly sensitive to the assumption on the mass density profile of the lensing galaxies, and various $H(z)$ models could be degenerate during the integration in the $D_{\Delta t}$ calibrator.  Fortunately, there is another high-redshift calibrator for IDL from the cosmic chronometer (CC) data~\cite{Jimenez:2001gg}, which directly measures the Hubble expansion rate at high redshifts by  $H(z)=-\mathrm{d}z/\mathrm{d}t/(1+z)$ without inputting any presumption on the cosmological model.

Second, the traditional IDL method usually fits to a cosmological model with Taylor expansion of $H(z)$ in redshifts $z$ or $y\equiv1-a=z/(1+z)$~\cite{Cattoen:2007sk}. As shown in Fig.~\ref{fig:BAO}, the Taylor expansions in $z$ or $y$ even up to the fifth order still largely distort the model that they are trying to approximate in the first place even for the $\Lambda$CDM model, hence fitting to the traditional Taylor expansion in $z$ or $y$ would be misleading for the BAO data considered to date. Fortunately, there is a global parametrization based on the cosmic age (PAge)~\cite{Huang:2020mub,Luo:2020ufj} that are faithful in reproducing a large class of late-time models in a wide redshift range with a high accuracy. See also~\cite{Huang:2021aku} for a more accurate parametrization based on cosmic age (MAPAge). 
Therefore, we improve the traditional IDL method by fitting the PAge model to CC+BAO+SNe data with a free $M_B$ prior~\cite{Cai:2021weh} (see also~\cite{Zhang:2020uan,Gomez-Valent:2021hda} for similar proposals and applications of this improved IDL with CC+BAO+SNe), and found no appealing evidence to go beyond the $\Lambda$CDM model at late time parametrized by the PAge model. This further strengths the usual no-go argument from IDL on the late-time physics.

On the other hand, the matter perturbation data is known to result in another tension called $S_8$ tension (see, e.g.~\cite{DiValentino:2020vvd,Perivolaropoulos:2021jda} and references therein), where $S_8=\sigma_8(\Omega_\mathrm{m}/0.3)^{0.5}$ is defined as a reflection of the degeneracy between the present-day matter density fraction $\Omega_\mathrm{m}$ and the root mean square (rms) of the matter density fluctuation $\sigma_8$ within a spherical top-hat window of comoving radius $8h^{-1}$ Mpc at the present day. The $S_8$ tension emerges for the $\Lambda$CDM model  predicting a stronger growth of matter perturbations from CMB data than those constrained by weak lensing (WL), cluster abundance (CA), and redshift space distortion (RSD). For example, Planck 2018 (TT,TE,EE+lowE) gives rise to $S_8=0.834\pm0.016$ ~\cite{Planck:2018vyg}, which is consistent with ACT+WMAP constraint on $S_8=0.840\pm0.030$~\cite{ACT:2020gnv}, but in a mild tension with the most recent WL constraint $S_8=0.759_{-0.021}^{+0.024}$ from the Kilo Degree Survey-1000 (KiDS-1000)~\cite{KiDS:2020suj} or even a severe tension with the most recent CA constraint (joining with WL) $S_8=0.65\pm0.04$ from the Dark Energy Survey Year 1 (DES-Y1)~\cite{DES:2020ahh}. The RSD growth data measures a different quantity called $f\sigma_8(z)$ (defined later below), which could lead to a constraint on $S_8=0.700_{-0.037}^{+0.038}$ from the most recent RSD selected dataset~\cite{Benisty:2020kdt}. The large variance in the inferred values of $S_8$ parameter makes us wander if there is really a $S_8$ tension at all~\cite{Nunes:2021ipq,Huang:2021tvo}.

Recently, an alternative view on the normalization of matter power spectrum~\cite{Sanchez:2020vvb} might shed light on the $S_8$ tension. The amplitude of the power spectrum $P(k)$ is usually normalized by $\sigma_8$ characterizing the rms linear perturbation variance in a sphere of radius $R=8 h^{-1}$ Mpc, which is controlled by both the dimensionless Hubble constant $h$ and the amplitude $A_s$ of the primordial scalar perturbations. However, $\sigma_8$ varies with $h$ not only from the change in the amplitude of $P(k)$ itself, but also from the change in the reference scale $R=8 h^{-1}$ Mpc. Thus, the use of the unit $h^{-1}$ Mpc in $R$ obscures the response of $P(k)$ to the changes in $h$. Therefore, Ref.~\cite{Sanchez:2020vvb}  proposes to use $\sigma_{12}$ defined as the rms linear perturbation variance in a sphere of radius $R=12$ Mpc to  eliminate the degeneracy of the reference scale $R=8 h^{-1}$ Mpc on the constraints on $h$. This new normalization not only brings the disagreement in WL constraints on $S_8=\sigma_8(\Omega_\mathrm{m}/0.3)^{0.5}$ between  Planck and DES into an excellent agreement in $S_{12}=\sigma_{12}(\omega_\mathrm{m}/0.14)^{0.4}$, but also suggests $f\sigma_{12}(z)$ as the most relevant quantity to describe RSD data. Therefore, in addition to the conventionally used $S_8$, we also adopt $S_{12}$ for data analysis.

In this paper, we continue our previous investigation on narrowing down the late-time solutions to the Hubble tension by including the matter perturbation growth data in addition to the background data made of the improved IDL from fitting the CC+BAO+SNe data to both PAge and MAPAge models. The outline is as follows: In Sec.~\ref{sec:model}, we depict our fitting models from two parametrizations based on the cosmic age. In Sec.~\ref{sec:data}, we describe the data we use for analysis. In Sec.~\ref{sec:results}, we summarize the results of data analysis. The last section~\ref{sec:con} is devoted for conclusions and discussions.

\section{Model}\label{sec:model}

\begin{figure*}
\centering
\includegraphics[width=0.9\textwidth]{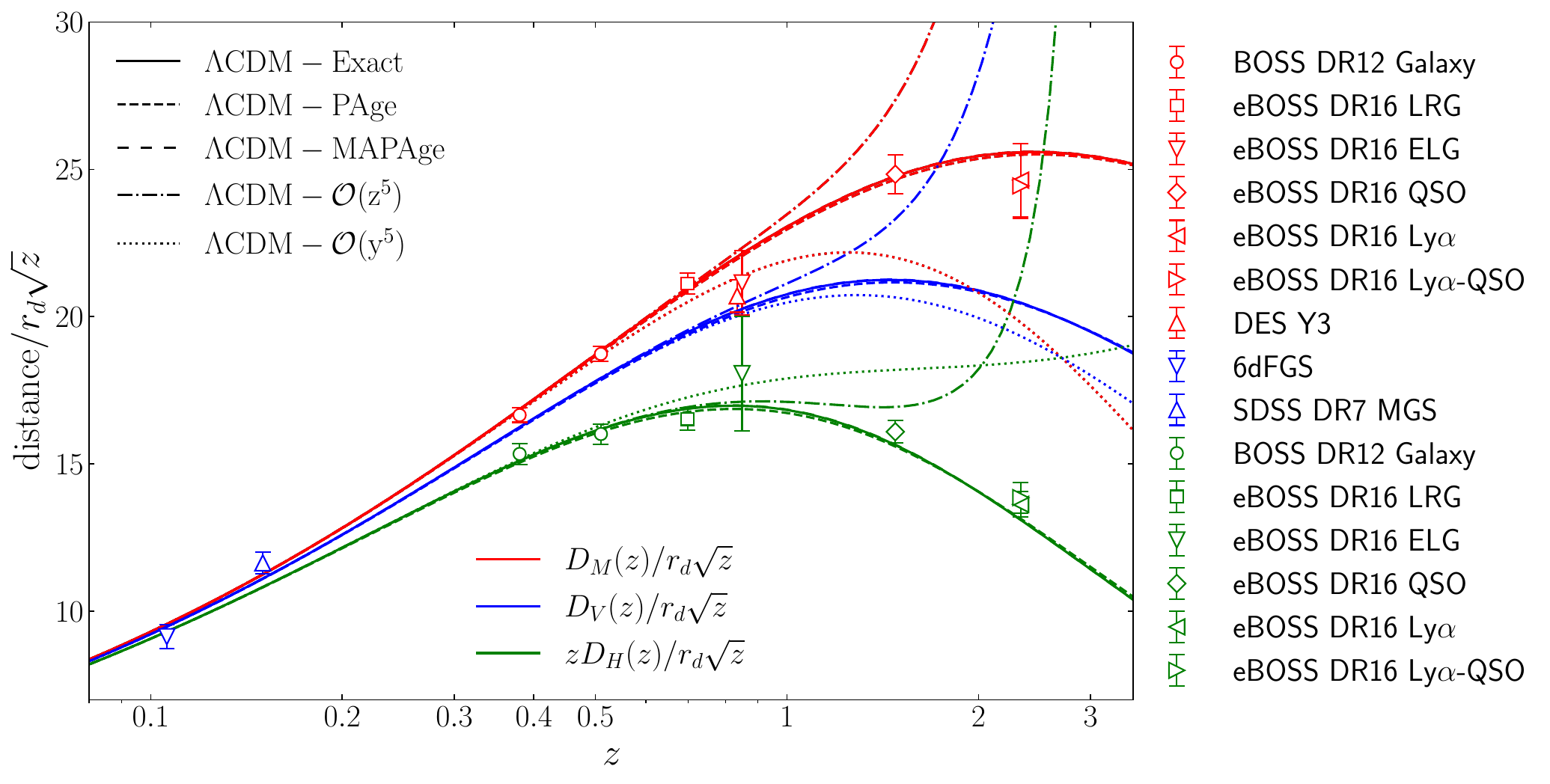}\\
\caption{The comparison of various BAO length scales ($D_M$ in red, $D_V$ in blue, and $D_H$ in green) among the exact $\Lambda$CDM model (solid),  the Taylor expansions of $\Lambda$CDM model in terms of the redshift $z$ (dash-dotted) and redshift $y\equiv1-a=z/(1+z)$ (dotted), and the PAge presentation of $\Lambda$CDM model (dash) with respect to the BAO data we use. The fiducial cosmology is assumed with $\Omega_m = 0.3156$, $H_0=67.27$ km/s/Mpc from Planck 2018 ~\cite{Planck:2018vyg}.}
\label{fig:BAO}
\end{figure*}

It is well-known for the $\Lambda$CDM model that the matter dominated era (9 Gyr), compared to the radiation dominated era (60 kyr), contributes to most of the cosmic age (13.7 Gyr) of our Universe, while the rest of period (4.7 Gyr) spans over the dark energy dominated era. This fact inspires Refs.  ~\cite{Huang:2020mub,Luo:2020ufj} to propose a global parametrization based on the cosmic age (PAge) for the late-time models  by expanding the product $Ht$ of the Hubble parameter $H(t)$ and the cosmic time $t$ to the quadratic order in $t$, namely,
\begin{align}\label{eq:PAge}
\frac{H}{H_0}=1+\frac23\left(1-\eta\frac{H_0t}{p_\mathrm{age}}\right)\left(\frac{1}{H_0t}-\frac{1}{p_\mathrm{age}}\right),
\end{align}
where the two free parameters $p_\mathrm{age}$ and $\eta$ characterize the deviation from the matter dominance combination $Ht=2/3$ when the dark energy takes over the background expansion. Here the short period spent by the radiation dominated era is neglected. Note that although $H_0$ explicitly appears at the right hand side of~\eqref{eq:PAge} in the combination $H_0t$, directly solving~\eqref{eq:PAge} for the combination $H_0t$ after replacing $H$ with $H(z)=-\mathrm{d}z/\mathrm{d}t/(1+z)$ gives rise to 
\begin{align}
1+z=\left(\frac{p_\mathrm{age}}{H_0t}\right)^\frac23e^{\frac13\left(1-\frac{H_0t}{p_\mathrm{age}}\right)\left(3p_\mathrm{age}+\eta\frac{H_0t}{p_\mathrm{age}}-\eta-2\right)},
\end{align}
where the combination $H_0t$ can now be solved as a function of $z$, $p_\mathrm{age}$ and $\eta$. Therefore, the dimensionless Hubble parameter $E(z)\equiv H(z)/H_0$ does not rely on $H_0$ but only the PAge parameters $p_\mathrm{age}$ and $\eta$. 

To see how a specific model is represented in the PAge parameter space $(p_\mathrm{age}, \eta)$, one first defines the PAge parameter $p_\mathrm{age}=H_0t_0$ as the product of current Hubble constant $H_0$ and cosmic age $t_0$, and then the matching of the time derivative of~\eqref{eq:PAge} to the present-day value $q_0$ of the deceleration parameter  $q(t)\equiv-\ddot{a}a/\dot{a}^2$ gives rise to a relation
\begin{align}\label{eq:eta}
\eta=1-\frac32p_\mathrm{age}^2(1+q_0).
\end{align}
For $\Lambda$CDM model with late-time parametrization $E(a)=\sqrt{\Omega_\mathrm{m}a^{-3}+1-\Omega_\mathrm{m}}$, the decelaration parameter is given by $q_0=-1+\frac32\Omega_\mathrm{m}$ and the current age of our Universe is given by
\begin{align}
t_0=\int_0^1\frac{\mathrm{d}a}{aH(a)}=\frac{9.77788\,\mathrm{Gyr}}{3h\sqrt{1-\Omega_\mathrm{m}}}\ln\frac{1+\sqrt{1-\Omega_\mathrm{m}}}{1-\sqrt{1-\Omega_\mathrm{m}}},
\end{align}
therefore, the PAge parameters $p_\mathrm{age}\equiv H_0t_0=0.9641$, $\eta\equiv1-\frac32p_\mathrm{age}^2(1+q_0)=0.3726$  for a fiducial cosmology with $\Omega_\mathrm{m}=0.3$ and $H_0=70$ km/s/Mpc. However, for a general late-time model, the relations $q_0(\Omega_\mathrm{m})$ and $t_0(\Omega_\mathrm{m})$ are arbitrary, hence the PAge parameters $\eta$ and $p_\mathrm{age}$ can be treated as independent parameters, which represent the deviation from the $\Lambda$CDM model by adding the duration of the dark-energy dominated era to the matter Universe differently from the $\Lambda$CDM model.

Note that the PAge representation of, for example,  the $\Lambda$CDM model, serves  as a much better approximation than the usual Taylor expansions in redshifts $z$ and $y$ even up to the fifth order at redshift $z\gtrsim1$ as shown in Fig.~\ref{fig:BAO}. 
Note also that different late-time models could be degenerated at the same point in the PAge parameter space $(p_\mathrm{age}, \eta)$. 
Therefore, the PAge approximation serves as a faithful and compact representation of the late-time models valid up to high redshift.
With the PAge approximation in hand, one can directly put cosmological constraints on the PAge parameters  $p_\mathrm{age}$ and $\eta$, and then map various specific late-time models onto the PAge parameter space  $(p_\mathrm{age}, \eta)$ without needing data analysis for these models one by one anymore. See~\cite{Cai:2021weh} for more details on applying this PAge model for Hubble-tension model-selections.

\begin{figure}
\centering
\includegraphics[width=0.4\textwidth]{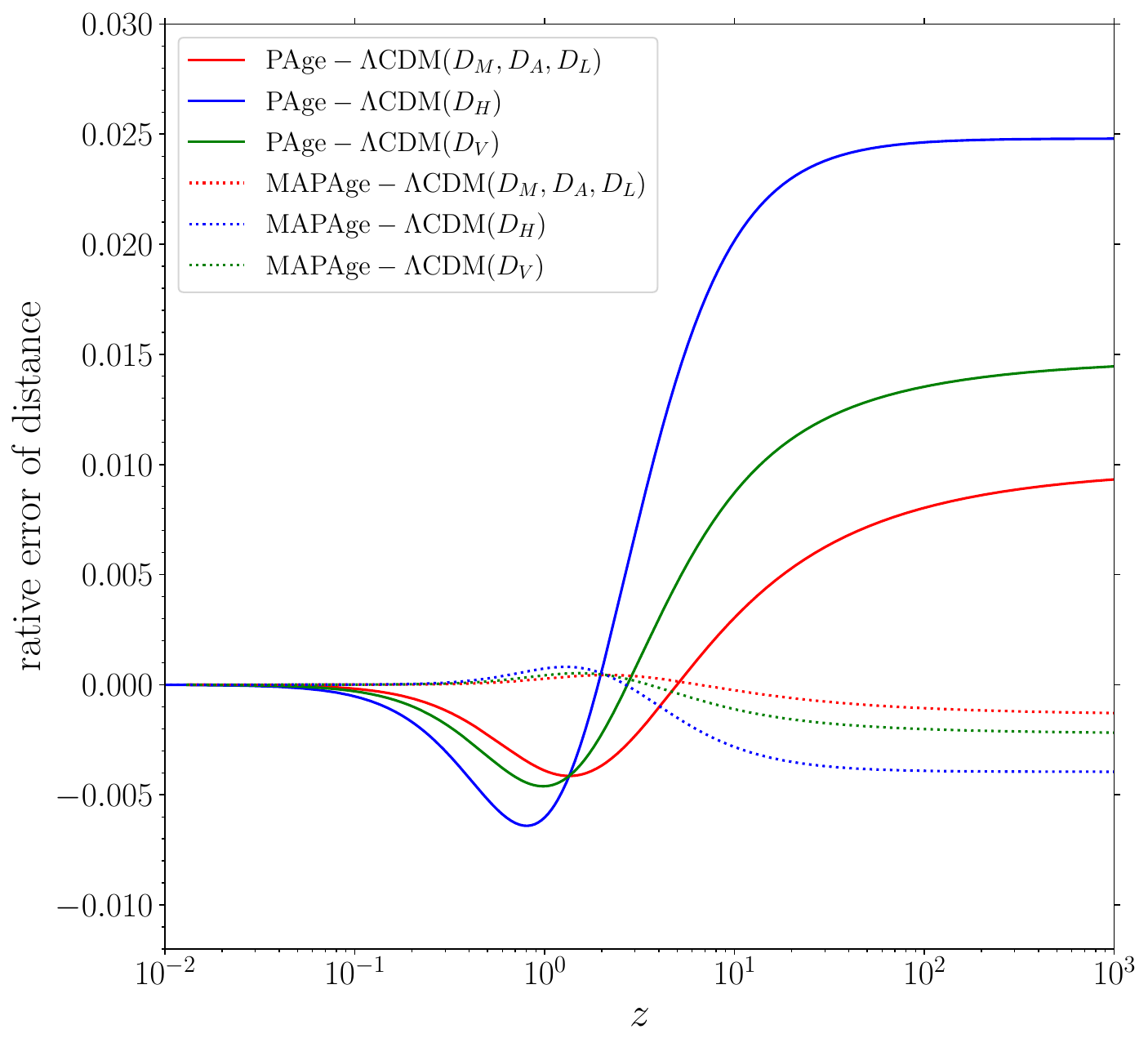}\\
\caption{The relative errors for various BAO length scales with PAge/MAPAge representation for the $\Lambda$CDM model with respect to the exact $\Lambda$CDM model in the same fiducial cosmology as Fig.~\ref{fig:BAO}.} 
\label{fig:comparison}
\end{figure}

Similar to the quadratic expansion of $Ht$ in $t$ where the PAge parameters $p_\mathrm{age}$ and $\eta$ capture the distortion to the cosmic age of the dust Universe from the late-time deceleration parameter, Ref.~\cite{Huang:2021aku} further proposed a more accurate parametrization based on the cosmic age (MAPAge) at the cubic order in $t$,
\begin{align}\label{eq:MAPAge}
\frac{H}{H_0}=1+\frac23&\left[1-(\eta+\eta_2)\frac{H_0t}{p_\mathrm{age}}+\eta_2\left(\frac{H_0t}{p_\mathrm{age}}\right)^2\right]\nonumber\\
&\times\left(\frac{1}{H_0t}-\frac{1}{p_\mathrm{age}}\right),
\end{align}
where the new PAge parameter $\eta_2$ further captures the original jerk parameter $j(t)=(\dddot{a}/a)/(\dot{a}/a)^3$ at, for example, the present day  by
\begin{align}
\eta_2=1-\frac34p_\mathrm{age}^3(2+j_0+3q_0).
\end{align}
The advantage of the MAPAge approximation over the PAge approximation is to gain more flexibility to characterize and distinguish the late-time models in a more accurate manner. See Fig.~\ref{fig:comparison} for a comparison of the relative errors of various BAO length scales (defined later)
\begin{align}
\frac{D_i(\mathrm{PAge/MAPAge\,\,of\,\,}\Lambda\mathrm{CDM})-D_i(\Lambda\mathrm{CDM})}{D_i(\Lambda\mathrm{CDM})}
\end{align} between the PAge and MAPAge representations of the $\Lambda$CDM model with respect to the exact $\Lambda$CDM expression. For the redshift range probed by the most recent BAO and CC data (introduced later), the relative error of the PAge (MAPAge) representation of the $\Lambda$CDM model is below $0.5\%$ ($0.1\%$). Note that even up to the redshift as high as in the cosmic dawn, the relative error of the PAge (MAPAge) representation is still below $2.5\%$ ($0.5\%$), rendering the PAge/MAPAge approximation the only appealing model-independent parametrization when including the future BAO data from the cosmic dawn~\cite{Munoz:2019fkt}.

\section{Data and methodology}\label{sec:data}

The datasets we use include SNe Ia, BAO, CC, and RSD, the first three of which are combined into the improved IDL, and the last one is the representative data for matter perturbations. 

\subsection{SNe Ia data}\label{subsec:SN}

For the SNe Ia data, we use the Pantheon sample~\cite{Scolnic:2017caz} with 1048 SNe Ia in the redshift range $0.01<z<2.3$, which records the apparent $B$-band magnitude $m_B(z)$ defined theoretically by
\begin{align}\label{eq:mB}
m_B(z)=M_B+5\lg\frac{D_L(z)}{10\,\mathrm{pc}}\equiv a_B+5\lg d_L(z)
\end{align}
with $a_B\equiv5\lg 10^3c/(\mathrm{km\cdot s^{-1}})-5\lg h$, where $H_0\equiv 100h$ km/s/Mpc and the luminosity distance $D_L(z)$, after factorized out the $H_0$-dependent part into the absolute $B$-band magnitude $M_B$ of SNe Ia, is computed by
\begin{align}\label{eq:dL}
d_L(z)\equiv\frac{D_L(z)}{c/H_0}=(1+z_\mathrm{hel})\int_0^{z_\mathrm{cmb}}\frac{\mathrm{d}z'}{E(z')}
\end{align}
for a cosmological model with the dimensionless Hubble parameter $E(z)\equiv H(z)/H_0$. Here $z_\mathrm{hel}$ is the heliocentric redshift, and $z_\mathrm{cmb}$ is the redshift in the cosmological rest frame, which corrects for the fact that the factor $1+z$ in the definition of $D_L$ should be calculated using the apparent redshift due
entirely to the loss of photon energy caused by redshift. Note that $M_B$ will be left as a free parameter during the data analysis due to the $M_B$ tension we mentioned in the introduction.

We consider the full covariance matrix for the Pantheon sample~\cite{Scolnic:2017caz},
\begin{equation}
    \mathrm{C}_{ij} = \mathrm{D}_{\mathrm{stat},ij} + \mathrm{C}_{\mathrm{sys},ij},
\end{equation}
where the statistic errors $\mathrm{D}_{\mathrm{stat},ij}$ is a diagonal matrix containing photometric error, mass step correction, peculiar velocity uncertainty, distance bias correction, stochastic gravitational lensing, redshift measurement uncertainty in quadrature and intrinsic scatter, while the systematic covariance $\mathrm{C}_{\mathrm{sys},ij}$ has nondiagonal parts.

\subsection{CC data}\label{subsec:CC}

For the CC data, we adopt the wildly-used compilation (see, e.g.,~\cite{Vagnozzi:2020dfn}) including 31 measurements~\cite{Jimenez:2003iv,Simon:2004tf,Stern:2009ep,Moresco:2012jh,Zhang:2012mp,Moresco:2015cya,Moresco:2016mzx,Ratsimbazafy:2017vga} within $0.07<z<1.965$ as well as the most recent measurements at $z=0.75$~\cite{Borghi:2021rft} as also listed in Table~\ref{tab:CC} for your convenience. With the differential age method ~\cite{Jimenez:2001gg}, the Hubble expansion rate is directly estimated as
\begin{align}
H(z)=-\frac{1}{1+z}\frac{\mathrm{d}z}{\mathrm{d}t}
\end{align}
by measuring the age difference $\Delta t$ between two passively-evolving galaxies of the same formation time but separated by a small redshift interval $\Delta z$, which is independent of any cosmological model assumption. This could bought us an advantage for cosmological model selections since the main systematic dependence from the age estimation on the evolutionary stellar population synthesis model is of astrophysical origin and should affect identically on the cosmological models, hence the difference of cosmological late-time models should be insensitive to  the astrophysical systemtatics of CC data. Therefore, any detection of deviations of the PAge/MAPAge models from the $\Lambda$CDM model could be regarded as the smoking gun for the new physics at the late time.

We adopt the total covariance matrix for CC from the state-of-art estimations \cite{Moresco:2020fbm,Moresco:2022phi},
\begin{equation}
    \mathrm{Cov}_{ij}^\mathrm{tot} = \mathrm{Cov}_{ij}^\mathrm{stat} + \mathrm{Cov}_{ij}^\mathrm{syst},
\end{equation}
where $\mathrm{Cov}_{ij}^\mathrm{stat}$ accounts for statistic errors and the systematic part $\mathrm{Cov}_{ij}^\mathrm{syst}$ can be decomposed into
\begin{equation}
    \mathrm{Cov}_{ij}^\mathrm{syst} = \mathrm{Cov}_{ij}^\mathrm{met} + \mathrm{Cov}_{ij}^\mathrm{young} + \mathrm{Cov}_{ij}^\mathrm{model}.
\end{equation}
Here $\mathrm{Cov}_{ij}^\mathrm{met}$ is the uncertainty in estimating the stellar metallicity, $\mathrm{Cov}_{ij}^\mathrm{young}$ is considered due to the effect from an eventual residual young component in galaxy spectra, and the modeling uncertainty $\mathrm{Cov}_{ij}^\mathrm{model}$ 
is constructed from the uncertainty in the star formation history (SFH), $\mathrm{Cov}_{ij}^\mathrm{SFH}$, the uncertainty in the initial initial mass functions (IMF), $\mathrm{Cov}_{ij}^\mathrm{IMF}$, the uncertainty in the stellar library adopted, $\mathrm{Cov}_{ij}^\mathrm{st.lib.}$, and the uncertainty in the stellar population synthesis (SPS) model adopted, namely,
\begin{align}
\mathrm{Cov}_{ij}^\mathrm{model} = \mathrm{Cov}_{ij}^\mathrm{SFH} + \mathrm{Cov}_{ij}^\mathrm{IMF} + \mathrm{Cov}_{ij}^\mathrm{st.lib} + \mathrm{Cov}_{ij}^\mathrm{SPS}
\end{align}
We use the suggested combination $\mathrm{Cov}_{ij}^\mathrm{tot}=\mathrm{Cov}_{ij}^\mathrm{diag}+\mathrm{Cov}_{ij}^\mathrm{sps,ooo}+\mathrm{Cov}_{ij}^\mathrm{IMF}$~\footnote{\url{https://gitlab.com/mmoresco/CCcovariance/-/blob/master/examples/CC_covariance_components.ipynb}}, where the diagonal matrix includes contributions from $\mathrm{Cov}_{ij}^\mathrm{stat}$, $\mathrm{Cov}_{ij}^\mathrm{met}$, $\mathrm{Cov}_{ij}^\mathrm{young}$, and $\mathrm{Cov}_{ij}^\mathrm{SFH}$, while the SPS model adopts the ``odd one out'' estimate. Note that the suggested combination does not include $\mathrm{Cov}_{ij}^\mathrm{st.lib}$, which is considered for a more conservative estimation. For those data points not discussed in Ref.~\cite{Moresco:2020fbm} (see also Table 1 of~\cite{Moresco:2022phi}), only the diagonal part is included.

\begin{table}
\caption{CC data}
\begin{tabular}{|c|c|c|}
\hline
\hline
$z$ & $H(z)$ km/s/Mpc  & References \\
\hline
\hline
0.1 & $69 \pm 12$ &~\cite{Jimenez:2003iv,Stern:2009ep}  \\
0.17 & $83 \pm 8$ &~\cite{Simon:2004tf,Stern:2009ep} \\
0.27 & $77 \pm 14$ &~\cite{Simon:2004tf,Stern:2009ep} \\
0.4 & $95 \pm 17$ &~\cite{Simon:2004tf,Stern:2009ep}  \\
0.48 & $97 \pm 62$ &~\cite{Stern:2009ep}  \\
0.88 & $90 \pm 40$ &~\cite{Stern:2009ep} \\
0.9 & $117 \pm 23$ &~\cite{Simon:2004tf,Stern:2009ep}  \\
1.3 & $168 \pm 17$ &~\cite{Simon:2004tf,Stern:2009ep} \\
1.43 & $177 \pm 18$ &~\cite{Simon:2004tf,Stern:2009ep} \\
1.53 & $140 \pm 14$ &~\cite{Simon:2004tf,Stern:2009ep} \\
1.75 & $202 \pm 40$ &~\cite{Simon:2004tf,Stern:2009ep} \\
\hline
\hline
0.1791 & $75 \pm 4$ &~\cite{Moresco:2012jh} \\
0.1993 & $75 \pm 5$ &~\cite{Moresco:2012jh} \\
0.3519 & $83 \pm 14$ &~\cite{Moresco:2012jh} \\
0.5929 & $104 \pm 13$ &~\cite{Moresco:2012jh} \\
0.6797 & $92 \pm 8$ &~\cite{Moresco:2012jh} \\
0.7812 & $105 \pm 12$ &~\cite{Moresco:2012jh} \\
0.8754 & $125 \pm 17$ &~\cite{Moresco:2012jh} \\
1.037 & $154 \pm 20$ &~\cite{Moresco:2012jh} \\
\hline
\hline
0.07 & $69.0 \pm 19.6$ &~\cite{Zhang:2012mp} \\
0.12 & $68.6 \pm 26.2$ &~\cite{Zhang:2012mp} \\
0.20 & $72.9 \pm 29.6$ &~\cite{Zhang:2012mp} \\
0.28 & $88.8 \pm 36.6$ &~\cite{Zhang:2012mp} \\
\hline
\hline
1.363 & $160 \pm 33.6$ &~\cite{Moresco:2015cya} \\
1.965 & $186.5 \pm 50.4$ &~\cite{Moresco:2015cya} \\
\hline
\hline
0.3802 & $83.0 \pm 13.5$ &~\cite{Moresco:2016mzx} \\
0.4004 & $77.0 \pm 10.2$ &~\cite{Moresco:2016mzx} \\
0.4247 & $87.1 \pm 11.2$ &~\cite{Moresco:2016mzx} \\
0.4497 & $92.8 \pm 12.9$ &~\cite{Moresco:2016mzx} \\
0.4783 & $80.9 \pm 9$ &~\cite{Moresco:2016mzx} \\
\hline
\hline
0.47 & $89 \pm 49.6$ &~\cite{Ratsimbazafy:2017vga}\\
\hline
\hline
0.75 & $98.8 \pm 33.6$ &~\cite{Borghi:2021rft} \\
\hline
\hline
\end{tabular}
\label{tab:CC}
\end{table}

\begin{table}
\caption{BAO data}\label{tab:BAO}
\centering
\begin{tabular}{|c|c|c|r|}
\hline
\hline
$z_\mathrm{eff}$ & Measurement & Constraint & References \\
\hline
\hline
\multicolumn{4}{|r|}{6dFGS~\cite{Beutler:2011hx} }\\
\hline
$0.106$ & $r_d/D_V$ & $0.336 \pm 0.015$ &~\cite{Beutler:2011hx}\\
\hline
\hline
\multicolumn{4}{|r|}{SDSS DR7 MGS~\cite{Ross:2014qpa}}\\
\hline
$0.15$ & $D_V/r_d$ & $4.51 \pm 0.14$ &  ~\cite{eBOSS:2020yzd} \\
\hline
\hline
\multicolumn{4}{|r|}{SDSS BOSS DR12~\cite{Alam:2016hwk}}\\
\hline
$0.38$ & $D_M/r_d$ & $10.27\pm0.15$ & ~\cite{eBOSS:2020yzd} \\
$0.51$ & $D_M/r_d$ & $13.38\pm0.18$ & ~\cite{eBOSS:2020yzd} \\
$0.38$ & $D_H/r_d$ & $24.89\pm0.58$ &~\cite{eBOSS:2020yzd} \\
$0.51$ & $D_H/r_d$ & $22.43\pm0.48$ & ~\cite{eBOSS:2020yzd} \\
\hline
\hline
\multicolumn{4}{|r|}{eBOSS DR16 LRG ~\cite{Bautista:2020ahg,Gil-Marin:2020bct}}\\
\hline
$0.698$ & $D_M/r_d$ & $17.65 \pm 0.30$ &~\cite{eBOSS:2020yzd}\\
$0.698$ & $D_H/r_d$ & $19.78 \pm 0.46$ &~\cite{eBOSS:2020yzd}\\
\hline
\hline
\multicolumn{4}{|r|}{eBOSS DR16 ELG~\cite{deMattia:2020fkb,Tamone:2020qrl}}\\
\hline
$0.85$ & $D_H/r_d$ & $19.5\pm1.0$ &~\cite{eBOSS:2020yzd} \\
$0.85$ & $D_M/r_d$ & $19.6\pm2.1$ &~\cite{eBOSS:2020yzd} \\
\hline
\hline
\multicolumn{4}{|r|}{eBOSS DR16 QSO~\cite{Neveux:2020voa,Hou:2020rse}}\\
\hline
$1.48$ & $D_M/r_d$ & $30.21 \pm 0.79$ &~\cite{eBOSS:2020yzd} \\
$1.48$ & $D_H/r_d$ & $13.23 \pm 0.47$ &~\cite{eBOSS:2020yzd} \\
\hline
\hline
\multicolumn{4}{|r|}{eBOSS DR16 Ly$\alpha$~\cite{duMasdesBourboux:2020pck}}\\
\hline
$2.33$ & $D_M/r_d$ & $37.6 \pm 1.9$ &~\cite{eBOSS:2020yzd}\\
$2.33$ & $D_H/r_d$ & $8.93 \pm 0.28$ &~\cite{eBOSS:2020yzd}\\
\hline
\hline
\multicolumn{4}{|r|}{eBOSS DR16 Ly$\alpha$-QSO~\cite{duMasdesBourboux:2020pck}}\\
\hline
$2.33$ & $D_M/r_d$ & $37.3 \pm 1.7$ &~\cite{eBOSS:2020yzd}\\
$2.33$ & $D_H/r_d$ & $9.08 \pm 0.34$ &~\cite{eBOSS:2020yzd}\\
\hline
\hline
\multicolumn{4}{|r|}{DES Y3~\cite{DES:2021esc}}\\
\hline
$0.835$ & $D_M/r_d$ & $18.92 \pm 0.51$ &~\cite{DES:2021esc} \\
\hline
\hline
\end{tabular}
\end{table}

\begin{table*}
\caption{RSD data}\label{tab:RSD}
\centering
\begin{tabular}{|l|c|c|c|l|r|}
\hline
\hline
Index & Dataset & Redshift & $f\sigma_8(z)$ & Fiducial cosmology & References \\
\hline
\hline
\multicolumn{6}{|r|}{2017 dataset~\cite{Nesseris:2017vor}}\\
\hline
1 & 6dFGS+SnIa & 0.02 & $0.428\pm 0.0465$ & $(\Omega_\mathrm{m}, h,\sigma_8)=(0.3, 0.683, 0.8)$ &~\cite{Huterer:2016uyq} \\
2 & SnIa+IRAS & 0.02 & $0.398\pm0.065$ & $(\Omega_\mathrm{m},\Omega_k)=(0.3,0)$ &~\cite{2012ApJ...751L..30H,2012MNRAS.420..447T}\\
3 & 2MASS & 0.02 & $0.314\pm0.048$ & $(\Omega_\mathrm{m},\Omega_k)=(0.266,0)$ &~\cite{2012ApJ...751L..30H,Davis:2010sw}\\
4 & SDSS-veloc & 0.10 & $0.370\pm0.130$ & $(\Omega_\mathrm{m},\Omega_k)=(0.3,0)$ &~\cite{Feix:2015dla}\\
5 & SDSS-MGS & 0.15 & $0.490\pm0.145$ & $(\Omega_\mathrm{m}, h,\sigma_8)=(0.31, 0.67, 0.83)$ &~\cite{Howlett:2014opa}\\
6 & 2dFGRS & 0.17 & $0.510\pm0.060$ &  $(\Omega_\mathrm{m},\Omega_k)=(0.3,0)$ &~\cite{Song:2008qt}\\
7 & GAMA & 0.18 & $0.360\pm0.090$ &  $(\Omega_\mathrm{m},\Omega_k)=(0.27,0)$ &~\cite{Blake:2013nif}\\
8 & GAMA & 0.38 & $0.440\pm0.060$ &  $(\Omega_\mathrm{m},\Omega_k)=(0.27,0)$ &~\cite{Blake:2013nif}\\
9 & SDSS-LRG-200 & 0.25 & $0.3512\pm0.0583$ &  $(\Omega_\mathrm{m},\Omega_k)=(0.25,0)$ &~\cite{2012MNRAS.420.2102S}\\
10 & SDSS-LRG-200 & 0.37 & $0.4602\pm0.0378$ &  $(\Omega_\mathrm{m},\Omega_k)=(0.25,0)$ &~\cite{2012MNRAS.420.2102S}\\
11 & BOSS-LOWZ & 0.32 & $0.384\pm0.095$ &  $(\Omega_\mathrm{m},\Omega_k)=(0.274,0)$ &~\cite{Sanchez:2013tga}\\
12 & SDSS-CMASS & 0.59 & $0.488\pm0.060$ & $(\Omega_\mathrm{m}, h,\sigma_8)=(0.307, 0.678, 0.829)$ &~\cite{Chuang:2013wga}\\
13 & WiggleZ & 0.44 & $0.413\pm0.080$ & $(\Omega_\mathrm{m},h)=(0.27, 0.71)$ &~\cite{2012MNRAS.425..405B}\\
14 & WiggleZ & 0.60 & $0.390\pm0.063$ & $(\Omega_\mathrm{m},h)=(0.27, 0.71)$ &~\cite{2012MNRAS.425..405B}\\
15 & WiggleZ & 0.73 & $0.437\pm0.072$ & $(\Omega_\mathrm{m},h)=(0.27, 0.71)$ &~\cite{2012MNRAS.425..405B}\\
16 & Vipers PDR-2 & 0.60 & $0.550\pm0.120$ & $(\Omega_\mathrm{m},\Omega_\mathrm{b})=(0.3, 0.045)$  &~\cite{Pezzotta:2016gbo}\\
17 & Vipers PDR-2 & 0.86 & $0.400\pm0.110$ & $(\Omega_\mathrm{m},\Omega_\mathrm{b})=(0.3, 0.045)$  &~\cite{Pezzotta:2016gbo}\\
18 & FastSound & 1.40 & $0.482\pm0.116$ &  $(\Omega_\mathrm{m},\Omega_k)=(0.27,0)$ &~\cite{Okumura:2015lvp}\\
\hline
\hline
\multicolumn{6}{|r|}{2018 dataset~\cite{Sagredo:2018ahx}}\\
\hline
1-18 & \multicolumn{5}{|l|}{the same as 2017 dataset 1-18}\\
\hline
19 & SDSS-IV eBOSS DR14 QSO & $0.978$ & $0.379\pm0.176$ & $\Omega_\mathrm{m}=0.31$ &~\cite{Zhao:2018gvb}\\
20 & SDSS-IV eBOSS DR14 QSO  & $1.23$ & $0.385\pm0.099$ & $\Omega_\mathrm{m}=0.31$ &~\cite{Zhao:2018gvb}\\
21 &  SDSS-IV eBOSS DR14 QSO & $1.526$ & $0.342\pm0.070$ & $\Omega_\mathrm{m}=0.31$ &~\cite{Zhao:2018gvb}\\
22 & SDSS-IV eBOSS DR14 QSO & $1.944$ & $0.364\pm0.106$ & $\Omega_\mathrm{m}=0.31$ &~\cite{Zhao:2018gvb}\\
\hline
\hline
\multicolumn{6}{|r|}{2021 dataset~\cite{Huang:2021tvo}}\\
\hline
1 & 2MTF & 0.001 & $0.505\pm 0.085$ & $(\Omega_\mathrm{m}, \sigma_8)=(0.312,0.815)$ &~\cite{Howlett:2017asq} \\
2 & ALFALFA & 0.013 & $0.46\pm 0.06$ & $(\Omega_\mathrm{m}, \sigma_8)=(0.315,0.8)$ &~\cite{Avila:2021dqv}\\
\hline
3-14 & \multicolumn{5}{|l|}{the same as 2017 dataset 1-3, 6-8, 13-18}\\
\hline
15 & SDSS DR7 MGS & 0.15 & $0.53\pm0.16$ & $(\Omega_\mathrm{m}, \sigma_8)=(0.31, 0.83)$ &~\cite{eBOSS:2020yzd} \\
16 & SDSS BOSS DR12 & 0.38 & $0.497\pm0.045$ & $(\Omega_\mathrm{m}, \sigma_8)=(0.31, 0.8)$ &~\cite{eBOSS:2020yzd} \\
17 & SDSS BOSS DR12 & 0.51 & $0.459\pm0.038$ & $(\Omega_\mathrm{m}, \sigma_8)=(0.31, 0.8)$ &~\cite{eBOSS:2020yzd} \\
18 & SDSS eBOSS DR16 LRG & 0.70 & $0.473\pm0.041$ & $(\Omega_\mathrm{m}, \sigma_8)=(0.31, 0.8)$ &~\cite{eBOSS:2020yzd} \\
19 & SDSS eBOSS DR16 ELG & 0.85 & $0.315\pm0.095$ & $(\Omega_\mathrm{m}, \sigma_8)=(0.31, 0.8)$ &~\cite{eBOSS:2020yzd} \\
20 & SDSS eBOSS DR16 QSO & 1.48 & $0.462\pm0.045$ & $(\Omega_\mathrm{m}, \sigma_8)=(0.31, 0.8)$ &~\cite{eBOSS:2020yzd} \\
\hline
\hline
\end{tabular}
\end{table*}

\subsection{BAO data}\label{subsec:BAO}

For BAO data, we adopt the state-of-the-art compilation from the final release of Sloan Digital Sky Survey (SDSS)-IV for the extended Baryon Oscillation Spectroscopic Survey (eBOSS) data release 16 (DR16) BAO+RSD measurements~\cite{eBOSS:2020yzd} as well as the Six-degree Field Galaxy Survey (6dFGS) measurement~\cite{Beutler:2011hx} and the most recent measurement from the Dark Energy Survey Year 3 (DES Y3)~\cite{DES:2021esc} as also listed in Table~\ref{tab:BAO} and Fig.~\ref{fig:BAO} for your convenience. The characteristic BAO length scales are measured in a dimensionless manner with respect to some fiducial cosmology as
\begin{align}
D_H(z)/r_d&=\frac{c}{H(z)r_d},\\
D_M(z)/r_d&=\frac{D_L(z)}{(1+z)r_d}=(1+z)D_A(z)/r_d,\\
D_V(z)/r_d&=\left[z D_M(z)^2D_H(z)\right]^{1/3}/r_d,
\end{align}
where the sound horizon at the drag epoch $r_d$ will be left as a free parameter during the data analysis due to the $r_d$ tension we mentioned in the introduction.
Note that the 6dFGS measurement is included since it is the lowest redshift BAO measurement ever made as a competitive and independent alternative to the other low-redshift BAO measurement from SDSS DR7 Main Galaxy Sample (MGS)~\cite{Ross:2014qpa}. Note also that the state-of-the-art compilation from the eBOSS DR16~\cite{eBOSS:2020yzd}  re-analyses all four generations of SDSS data, for example, SDSS DR7 MGS~\cite{Ross:2014qpa}, SDSS-III BOSS DR12~\cite{Alam:2016hwk}, SDSS-IV eBOSS DR16 Luminous Red Galaxies (LRG)~\cite{Bautista:2020ahg,Gil-Marin:2020bct}, SDSS-IV eBOSS DR16 Emission Line Galaxies (ELG) ~\cite{deMattia:2020fkb,Tamone:2020qrl}, SDSS-IV eBOSS DR16 Quasar Sample (QSO)~\cite{Neveux:2020voa,Hou:2020rse}, SDSS-IV eBOSS DR16 Ly$\alpha$~\cite{duMasdesBourboux:2020pck}, and then incorporates the systematic errors and consensus estimates into the covariance matrices to obtain the combined BAO+RSD measurements with inclusions of both Alcock-Paczynski (AP) effect~\cite{Alcock:1979mp} and reconstruction procedure~\cite{Eisenstein:2006nk}. Although deviated from the Planck $\Lambda$CDM prediction by $2.27\sigma$, the most recent BAO measurement from the DES Y3~\cite{DES:2021esc} is still included since it is the most precise measurement in the redshift range $0.6\lesssim z\lesssim 1.1$ to date.
For the final release of BAO+RSD measurements from SDSS-IV eBOSS DR16 (including the reanalysis of BOSS DR12), we adopt the same full covariance matrixes from the public SDSS svn repository ~\footnote{\url{https://svn.sdss.org/public/data/eboss/DR16cosmo/tags/v1_0_0/likelihoods/BAO-plus/}} as used in Ref.~\cite{eBOSS:2020yzd}.

\subsection{RSD data}\label{subsec:RSD}

For RSD data, we test for three $f\sigma_8$ compilations~\cite{Nesseris:2017vor,Sagredo:2018ahx,Huang:2021tvo} (dubbed as RSD 2017, 2018 and 2021 datasets) as also listed in Table~\ref{tab:RSD} for your convenience. The matter density perturbations $\delta(\mathbf{x}, t)\equiv\rho_\mathrm{m}(\mathbf{x},t)/\bar{\rho}_\mathrm{m}(t)-1$ defined by the excess of the matter density $\rho_\mathrm{m}(\mathbf{x},t)$ at the comoving coordinate $\mathbf{x}$ with respect to the mean matter density at the same cosmic time $t$ grows scale-independently from some initial-time density contrast by $\delta(\mathbf{x},t)=D(t)\delta(\mathbf{x},t_i)$ in the first-order perturbation theory, where the growth factor $D(a(t))$ obeys
\begin{align}\label{eq:Da}
D''(a)+\left(\frac{3}{a}+\frac{E'(a)}{E(a)}\right)D'(a)-\frac{3\Omega_\mathrm{m}D(a)}{2a^5E(a)^2}=0
\end{align}
with initial conditions $D(a_i)=1$ and $D'(a_i)=1/a_i$. In practice, the initial scale factor can be chosen as $a_i=0.001$ deep into the matter dominated era. The linear growth rate is then defined as the logarithmic derivative of the
growth factor with respect to the scale factor,
\begin{align}
f(a)=\frac{\mathrm{d}\ln D}{\mathrm{d}\ln a},
\end{align}
and the RSD measurements provide constraints on the bias-free combination $f\sigma_8\equiv f(a)\sigma_8(a)$ estimated by
\begin{align}
f(a)\sigma_8(a)=a\sigma_8(a_i)D'(a)=\sigma_8\frac{aD'(a)}{D(a=1)},
\end{align}
where $\sigma_8(a)$, the amplitude of linear matter perturbations on a comoving scale of $8h^{-1}$ Mpc at a redshift $z=1/a-1$, is related to its present-day value $\sigma_8$ by $\sigma_8(a)=D(a)\sigma_8(a_i)=D(a)\sigma_8/D(a=1)$. For $\Lambda$CDM model, the Eq.~\eqref{eq:Da} can be solved approximately by $f(z)\approx\Omega_\mathrm{m}(z)^{0.55}$~\cite{Linder:2007hg}. The $S_8$ parameter is therefore defined as $S_8=\sigma_8(\Omega_\mathrm{m}/0.3)^{0.5}$. It is worth noting that, to compare with the RSD measurements, the model estimation on $f\sigma_8$ should be corrected as~\cite{Macaulay:2013swa,Alam:2015rsa}
\begin{align}
(f\sigma)_8^\mathrm{corrected}=\frac{H^\mathrm{model}(z)D_A^\mathrm{model}(z)}{H^\mathrm{fiducial}(z)D_A^\mathrm{fiducial}(z)}\times(f\sigma)_8^\mathrm{model}
\end{align}
for the AP effect~\cite{Alcock:1979mp} that biases the $f\sigma_8$  measurements. 

As for the RSD $f\sigma_{12}$-data, there is yet no direct measurement (see, however,~\cite{Semenaite:2021aen} for the $\sigma_{12}$ measurement from the full shape BOSS+eBOSS QSO samples). We therefore adopt an indirect approach to reuse the existing RSD $f\sigma_8$-data by fixing $h=0.67$ during data analysis so that the constrained rms linear perturbation variance $\sigma_8$ within $8h^{-1}$ Mpc is in fact equal to the same matter perturbation $\sigma_{12}$ within $12$ Mpc. The reduced $S_{12}$ parameter is therefore estimated by $S_{12}=\sigma_{12}(\Omega_\mathrm{m}h^2/0.14)^{0.4}$ as defined in~\cite{Sanchez:2020vvb}.

For the RSD 2017 dataset (also known as the Gold 2017 compilation~\cite{Nesseris:2017vor}), note that it was specifically constructed and selected from a larger dataset to minimize the overlapping effect and maximize the independence of the data points. The Gold 2017 compilation has assumed that most of the data are not correlated but with the exception of the ones from WiggleZ given by~\cite{Blake:2012pj}. Although the survey redshift ranges are similar, the correlation between WiggleZ and BOSS-LOWZ/CMASS could be neglected since they are targeting at rather different types of galaxies, that is, the WiggleZ survey targets at emission-line galaxies while the BOSS-LOWZ/CMASS target at luminous red galaxies/massive galaxies. As an extension to the Gold 2017 compilation, the RSD 2018 dataset has adopted the covariance matrix from Ref.~\cite{Sagredo:2018ahx} for the four datapoints of the SDSDS-IV eBOSS DR14 QSO measurements. As a further update for the RSD 2017 and 2018 datasets, the RSD 2021 dataset has adopted the same covariance matrixes from the public SDSS svn repository ~\footnote{\url{https://svn.sdss.org/public/data/eboss/DR16cosmo/tags/v1_0_0/likelihoods/BAO-plus/}} as those used in Ref. \cite{eBOSS:2020yzd} for the final release BAO+RSD measurements.

\subsection{Data analysis}\label{subsec:method}

For data analysis, we fit the perturbative IDL data with inclusion of RSD 2018 and 2021 datasets (RSD 2018/2021+CC+BAO+SNe) to the $\Lambda$CDM, PAge and MAPAge models, respectively, with the Markov Chain Monte Carlo code \texttt{EMCEE}~\cite{Foreman-Mackey:2012any} to constrain the cosmological parameters with flat priors from the joint likelihood function $-2\ln\mathcal{L}=\chi^2_\mathrm{SN}+\chi^2_\mathrm{BAO}+\chi^2_\mathrm{CC}+\chi^2_\mathrm{RSD}$. For comparison, we also fit the same model to the background IDL data (CC+BAO+SNe) without $f\sigma_8$ data and also chop off the RSD part of the likelihood \footnote{See \url{https://svn.sdss.org/public/data/eboss/DR16cosmo/tags/v1_0_0/likelihoods/BAO-plus/}} for the BAO+RSD distance measurements~\cite{eBOSS:2020yzd}. For model comparison, we adopt the Akaike information criterion (AIC) $\mathrm{AIC} = 2k - 2 \ln(\mathcal{L}) $ and Bayesian information criterion (BIC)~\cite{Schwarz:1978tpv} $ \mathrm{BIC}= k \ln(n) - 2 \ln(\mathcal{L})$, where $n=1097, 1119, 1117$ are the numbers of data points in the CC+BAO+SNe, RSD 2018+CC+BAO+SNe, and RSD 2021+CC+BAO+SNe datasets, respectively, and $k=4; 5; 6$ are the numbers of free parameters of the $\Lambda$CDM, PAge, and MAPAge models in fitting to the CC+BAO+SNe dataset,  and $k=5; 7; 8$ to both RSD 2018/2021+CC+BAO+SNe datasets, respectively. For data analysis concerned with the reduced $S_{12}$ parameter, $k$ is also reduced by 1 due to the fixed $h=0.67$.

\begin{table*}
\small
\centering
\caption{Cosmological constraints from fitting CC+BAO+SNe to the $\Lambda$CDM, PAge, and MAPAge models, respectively. The AIC and BIC values are estimated with respect to the $\Lambda$CDM model.}
\begin{tabular}{c|c|c|c|c|c|c|c}
\hline
\hline
\multirow{2}*{Parameter}  & \multirow{2}*{Uniform prior} & \multicolumn{3}{|c}{CC+BAO+SNe} 
& \multicolumn{3}{|c}{CC+BAO+SNe with fixed $H_0$ for $S_{12}$} \\ 
\cline{3-8}
 & & $\Lambda$CDM  & PAge & MAPAge & $\Lambda$CDM & PAge & MAPAge  \\
\hline
$\Omega_m$ & $(0.01,1)$ 
& $0.297^{+0.014}_{-0.013}$  & --- & --- & $0.302 \pm 0.013$ & --- & --- \\
$p_\mathrm{age}$ & $(0.4,2.0)$ 
& ---  & $0.971^{+0.013}_{-0.012}$  & $0.987^{+0.020}_{-0.018}$ & ---  & $0.967 \pm 0.012 $  & $0.983^{+0.020}_{-0.018}$\\
$\eta$ & $(-2,2)$ 
& --- & $0.336^{+0.065}_{-0.067}$ & $0.524^{+0.168}_{-0.162}$ & ---  & $0.322^{+0.066}_{-0.067}$ &  $0.510^{+0.169}_{-0.163}$ \\
$\eta_2$ & $(-3,2)$ 
& --- & --- & $-0.615^{+0.475}_{-0.530}$  & --- & --- & $-0.616^{+0.484}_{-0.533}$ \\
$H_0$ & $(40,110)$ 
& $70.1 \pm 2.5$ & $69.9 \pm 2.5$ & $69.9 \pm 2.5$ & $67$ & $67$ & $67$ \\
$M_B$ & $(-20,-19)$ 
& $-19.348^{+0.074}_{-0.076}$ & $-19.348^{+0.074}_{-0.075}$ & $-19.361^{+0.074}_{-0.077}$ &  $-19.444 \pm 0.007$&  $-19.436 \pm 0.011$   &  $-19.449^{+0.015}_{-0.016}$ \\
$r_d$ & $(100,200)$ 
& $143.9^{+5.0}_{-4.7}$ & $143.7^{+5.0}_{-4.8}$ & $144.3^{+5.1}_{-4.8}$ & $150.1 \pm 1.5$  & $149.3^{+1.7}_{-1.6} $  & $150.0^{+1.8}_{-1.7}$ \\
\hline
$\chi^2/\mathrm{d.o.f}$ & --- & $0.9669$ & $0.9663$ & $0.9660$ & $0.9675$
& $0.9667$   & $0.9664$\\
\hline
$\Delta \mathrm{AIC}$ & --- & $0$ &  $0.3479$
& $ 1.0463$ & $0$ & $0.0790$  & $ 0.8153$  \\
$\Delta \mathrm{BIC}$ & --- & $0$ &  $5.3483$
& $11.0470$ & $0$ & $5.0793$ & $10.8160$ \\
\hline
\hline
\end{tabular}
\label{tab:background}
\end{table*}

\begin{table*}
\small
\centering
\caption{Cosmological constraints (in particular $S_8$) from fitting RSD 2018/2021+CC+BAO+SNe to the $\Lambda$CDM, PAge, and MAPAge models, respectively. The AIC and BIC values are estimated with respect to the $\Lambda$CDM model.}
\begin{tabular}{c|c|c|c|c|c|c|c}
\hline
\hline
\multirow{2}*{Parameter} 
& \multirow{2}*{Uniform prior} 
& \multicolumn{3}{|c}{RSD 2018+CC+BAO+SNe} 
& \multicolumn{3}{|c}{RSD 2021+CC+BAO+SNe} \\
\cline{3-8} & & $\Lambda$CDM & PAge & MAPAge & $\Lambda$CDM & PAge & MAPAge\\
\hline
$\Omega_m$ & $(0.01,1)$ 
& $0.294 \pm 0.013 $ & $0.218^{+0.103}_{-0.078}$ & $0.219^{+0.104}_{-0.079}$ 
& $0.288^{+0.013}_{-0.012}$ & $0.178^{+0.054}_{-0.046}$ & $0.177^{+0.053}_{-0.046}$ \\
\multirow{2}*{$p_\mathrm{age}$ }  &  {\tiny $(0.4,2.0)$ for PAge}
& \multirow{2}*{---}  & \multirow{2}*{$0.971^{+0.013}_{-0.012}$}  &\multirow{2}*{$0.988^{+0.020}_{-0.018}$ } 
& \multirow{2}*{---}  &\multirow{2}*{ $0.972^{+0.013}_{-0.012}$} 
& \multirow{2}*{$0.990^{+0.020}_{-0.018}$ } \\

& {\tiny $(0.4,1.5)$ for MAPAge}  & & & & & \\
$\eta$ & $(-2,2)$ 
& --- & $0.335^{+0.066}_{-0.069}$ & $0.526^{+0.168}_{-0.162}$ 
& --- & $0.325^{+0.064}_{-0.067}$ & $0.531^{+0.166}_{-0.160}$ \\
$\eta_2$ & $(-3,2)$ 
& --- & --- & $-0.621^{+0.479}_{-0.532}$ 
& --- & --- & $-0.667^{+0.484}_{-0.531}$ \\
$H_0$ & $(40,110)$ 
& $70.2 \pm 2.5$ & $69.8^{+2.5}_{-2.4}$ & $69.9 \pm 2.4$ 
& $70.6 \pm 2.4$ & $69.9 \pm 2.5$ & $69.9 \pm 2.5$ \\
$M_B$ & $(-20,-19)$ 
& $-19.345^{+0.075}_{-0.077} $ & $-19.350^{+0.074}_{-0.075}$ & $-19.361^{+0.074}_{-0.076}$ 
& $-19.336^{+0.072}_{-0.075}$ &  $-19.347^{+0.074}_{-0.078} $ & $-19.360^{+0.074}_{-0.077}$ \\
$r_d$ & $(100,200)$ 
& $143.9^{+5.1}_{-4.8}$ & $143.7^{+5.0}_{-4.7}$ & $144.3^{+5.1}_{-4.7}$ 
& $143.9^{+5.0}_{-4.7}$ & $143.7^{+5.2}_{-4.8}$ & $144.3^{+5.1}_{-4.7}$ \\
$\sigma_8$ & $(0.1,3.0)$ 
& $0.768^{+0.031}_{-0.030}$ & $0.896^{+0.248}_{-0.164}$ & $0.886^{+0.245}_{-0.160}$
& $0.829^{+0.027}_{-0.026}$ & $1.06^{+0.18}_{-0.13}$ & $1.05^{+0.18}_{-0.13}$ \\
$S_8$ & --- 
& $0.760^{+0.029}_{-0.028} $ & $0.768^{+0.032}_{-0.030}$ & $0.763^{+0.032}_{-0.030}$ 
& $0.811^{+0.027}_{-0.026} $ & $0.817 \pm 0.026$ & $0.812^{+0.027}_{-0.026}$ \\
\hline
$\chi^2/\mathrm{d.o.f}$ & --- 
& $0.9605$ & $0.9602$ & $0.9600$
& $0.9718$ & $0.9682$ & $0.9677$ \\
\hline
$\Delta \mathrm{AIC}$ & --- 
& $0$ & $1.7964$ & $2.5612$
& $0$ & $-1.8758$ & $-1.4719$ \\
$\Delta \mathrm{BIC}$ & ---
& $0$ & $11.8368$ & $17.6218$
& $0$ & $8.1610$ & $13.5834$ \\
\hline
\hline
\end{tabular}
\label{tab:S8}
\end{table*}

\begin{table*}
\small
\centering
\caption{Cosmological constraints (reducing to $S_{12}$) from fitting RSD 2018/2021+CC+BAO+SNe to the $\Lambda$CDM, PAge, and MAPAge models, respectively. The AIC and BIC values are estimated with respect to the $\Lambda$CDM model.}
\begin{tabular}{c|c|c|c|c|c|c|c}
\hline
\hline
\multirow{2}*{Parameter} 
& \multirow{2}*{Uniform prior} 
& \multicolumn{3}{|c}{RSD 2018+CC+BAO+SNe} 
& \multicolumn{3}{|c}{RSD 2021+CC+BAO+SNe} \\
\cline{3-8} & & $\Lambda$CDM & PAge & MAPAge & $\Lambda$CDM & PAge & MAPAge\\
\hline
$\Omega_m$ & $(0.01,1)$ 
& $0.300^{+0.013}_{-0.012}$ & $0.215^{+0.104}_{-0.081}$ & $0.214^{+0.101}_{-0.080}$ 
& $0.293 \pm 0.012$ & $0.175^{+0.054}_{-0.047}$ & $0.178^{+0.053}_{-0.046}$ \\
\multirow{2}*{$p_\mathrm{age}$ } & {\tiny $(0.4,2.0)$ for PAge} 
& \multirow{2}*{---} & \multirow{2}*{$0.967 \pm 0.012 $} &\multirow{2}*{$0.983^{+0.019}_{-0.018}$ } 
& \multirow{2}*{---}& \multirow{2}*{$0.967 \pm 0.012$} &\multirow{2}*{$0.985^{+0.019}_{-0.018}$}  \\
& {\tiny $(0.4,1.5)$ for MAPAge}  & & & & & \\
$\eta$ & $(-2,2)$ 
& --- & $0.322^{+0.065}_{-0.067}$ & $0.506^{+0.165}_{-0.161}$ 
& --- & $0.313^{+0.065}_{-0.067}$ & $0.511^{+0.168}_{-0.161}$ \\
$\eta_2$ & $(-2,2)$ 
& --- & --- & $-0.601^{+0.475}_{-0.522}$ 
& --- & --- & $-0.648^{+0.479}_{-0.538}$ \\
$M_B$ & $(-20,-19)$ 
& $-19.445 \pm 0.007 $ & $-19.436 \pm 0.011$ & $-19.449 \pm 0.015$
& $-19.448 \pm 0.007 $ & $-19.435 \pm 0.011$ & $-19.449 \pm 0.015 $ \\
$r_d$ & $(100,200)$ 
& $150.3^{+1.5}_{-1.4}$ & $149.3 \pm 1.6 $ & $149.9 \pm 1.7$ 
& $151.2 \pm 1.4$ & $149.3 \pm 1.6 $ & $150.1 \pm 1.7$ \\
$\sigma_{12}$ & $(0,3.0)$ 
& $0.762 \pm 0.030$ & $0.905^{+0.269}_{-0.169}$ & $0.901^{+0.266}_{-0.164}$ 
& $0.826 \pm 0.026$ & $1.07^{+0.19}_{-0.14}$ & $ 1.06^{+0.18}_{-0.13}$ \\
$S_{12}$ & --- 
& $0.750 \pm 0.028$ & $0.782^{+0.063}_{-0.046}$ & $0.777^{+0.062}_{-0.044}$ 
& $0.806 \pm 0.025$ & $0.851^{+0.041}_{-0.035}$ & $0.845^{+0.039}_{-0.034}$ \\
\hline
$\chi^2/\mathrm{d.o.f}$ & --- 
& $0.9612$ & $0.9606$ & $0.9604$ 
& $0.9729$ & $0.9686$ & $0.9681$ \\
\hline
$\Delta \mathrm{AIC}$ & --- 
& $0$ & $1.3736$ & $2.1712$ 
& $0$ & $ -2.7448$ & $-2.2614$ \\
$\Delta \mathrm{BIC}$ & --- 
& $0$ & $11.4140$ & $17.2318$ 
& $0$ & $7.2920$ & $12.7938$ \\
\hline
\hline
\end{tabular}
\label{tab:S12}   
\end{table*}

\section{Results}\label{sec:results}

\begin{figure}
\centering
\includegraphics[width=0.5\textwidth]{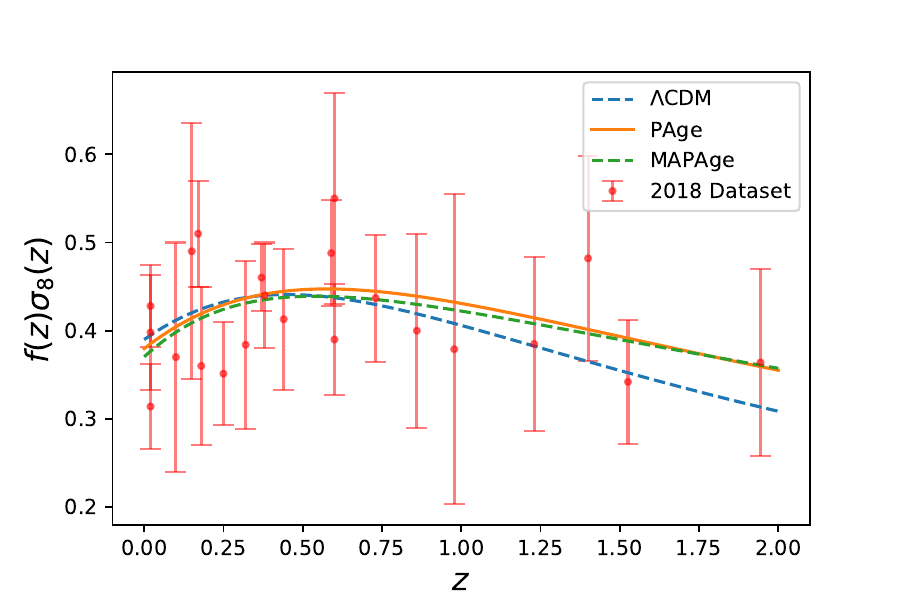}
\includegraphics[width=0.5\textwidth]{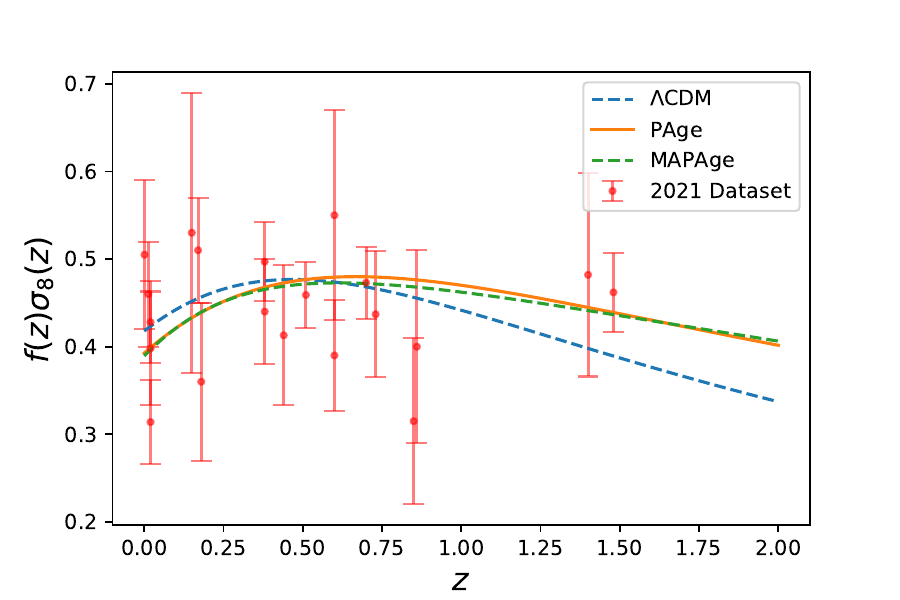}\\
\caption{The comparison between the best-fit models from $\Lambda$CDM (blue dashed curve), PAge (orange solid curve), and MAPAge (green dashed curve) with respect to the 2018 RSD dataset (top panel) and 2021 RSD dataset (bottom panel).}
\label{fig:RSDDataset}
\end{figure}

Cosmological constraints from fitting the perturbative IDL data (RSD 2018/2021+CC+BAO+SNe) to the $\Lambda$CDM, PAge, and MAPAge models, respectively, are summarized in Table~\ref{tab:S8} (including $S_8$ constraint) and Table~\ref{tab:S12} (with reducing constraint on $S_{12}$). For comparison, we also summarize the background cosmological constraints from fitting the background IDL data (CC+BAO+SNe) to the $\Lambda$CDM, PAge, and MAPAge models, respectively, in Table~\ref{tab:background}, where the constraints with fixed $h=0.67$ will be used for comparison to the constraints in Table~\ref{tab:S12} with RSD data and reduced $S_{12}$ parameter. The RSD datasets and the best fits of individual models are depicted in Fig.~\ref{fig:RSDDataset}. All the AIC/BIC values are estimated with respect to the $\Lambda$CDM model in the corresponding data fitting combinations. 

\begin{figure}
\centering
\includegraphics[width=0.4\textwidth]{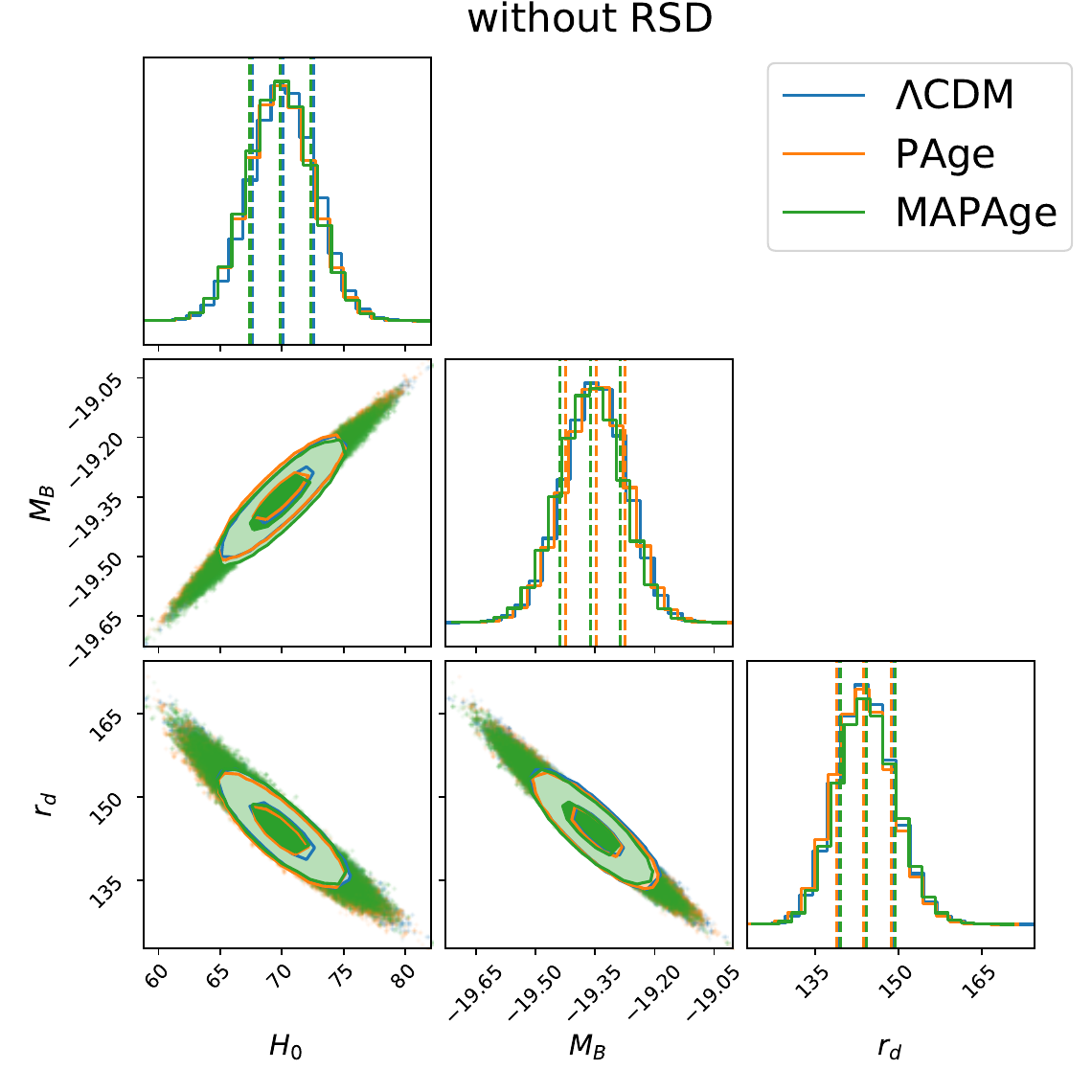}
\includegraphics[width=0.4\textwidth]{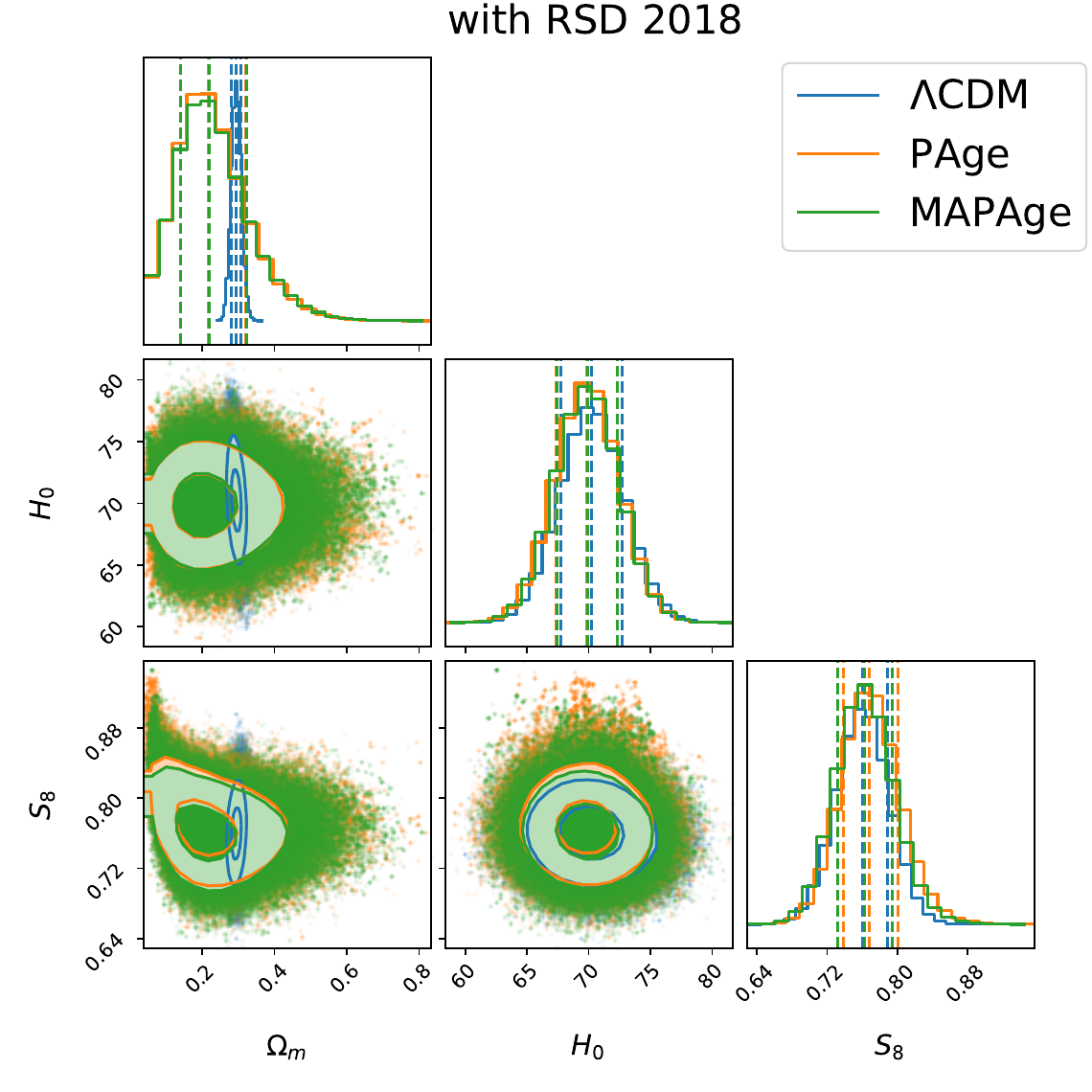}
\includegraphics[width=0.4\textwidth]{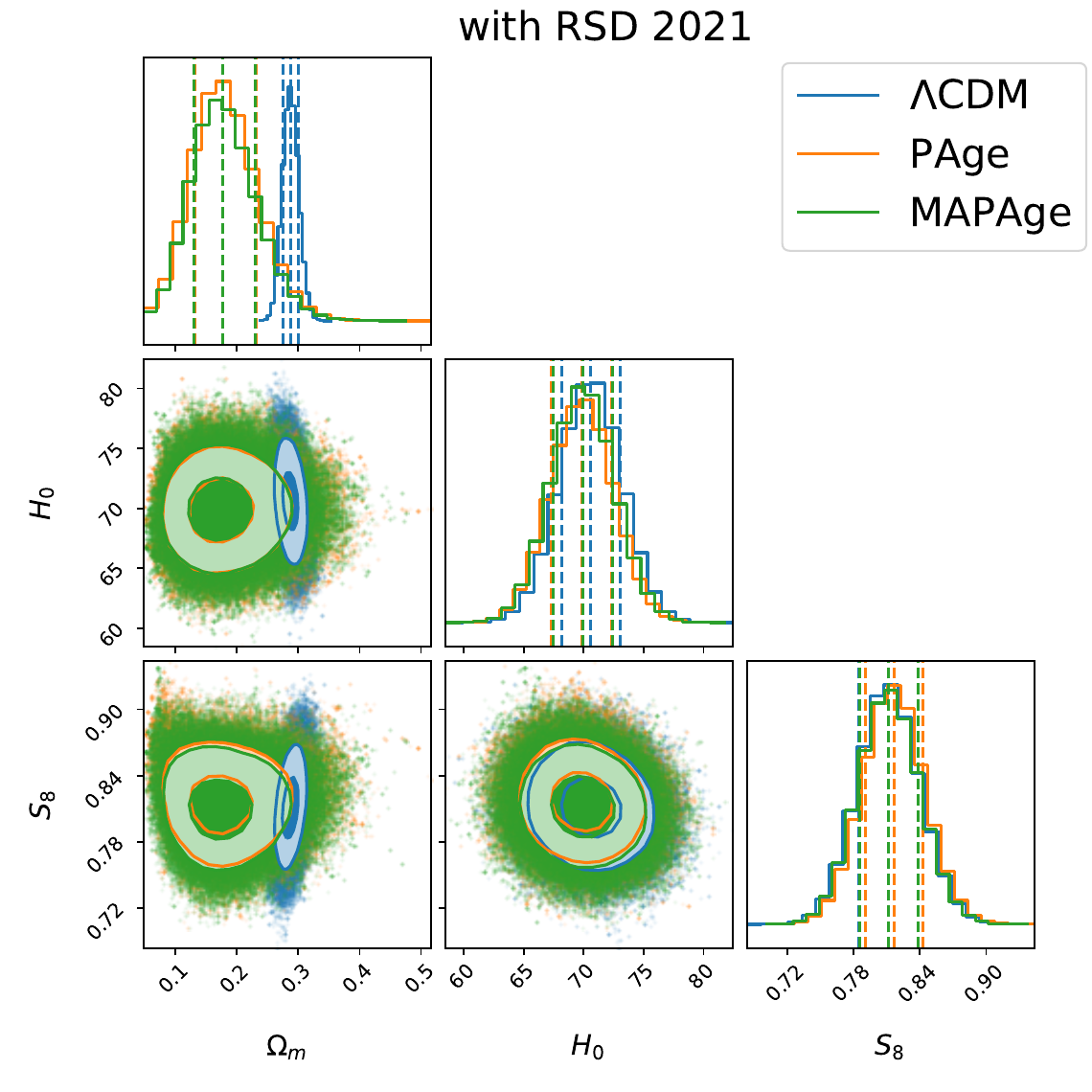}\\
\caption{Model comparison from fitting the $\Lambda$CDM (blue), PAge (orange), and MAPAge (green) models to the background IDL (CC+BAO+SNe) data (top) and perturbative IDL data (RSD+CC+BAO+SNe) with inclusions of RSD 2018 (medium) and RSD 2021 (bottom) data.}
\label{fig:modelcomparison}
\end{figure}

\begin{figure}
\centering
\includegraphics[width=0.4\textwidth]{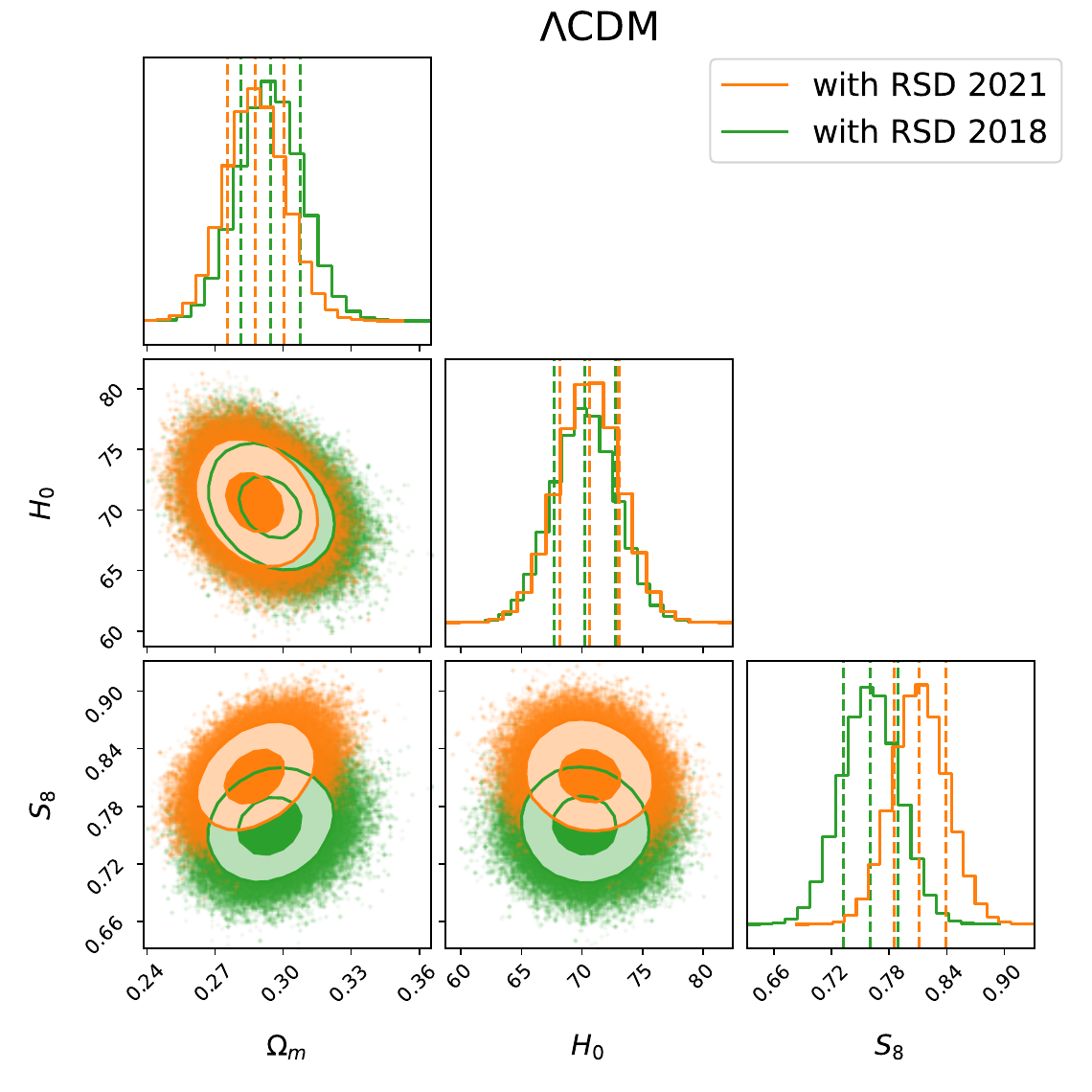}
\includegraphics[width=0.4\textwidth]{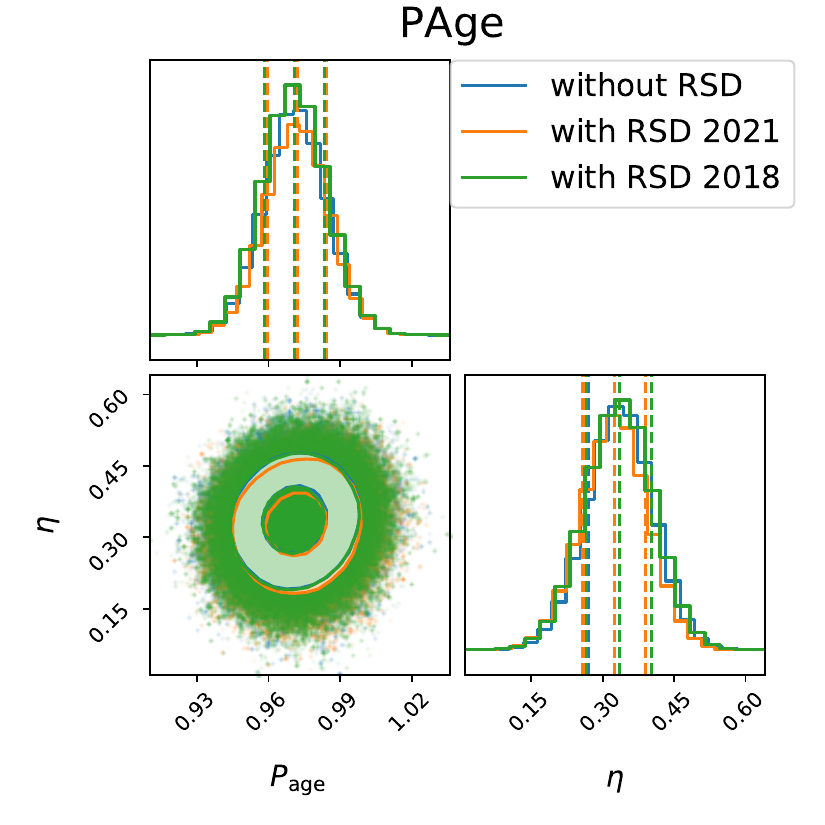}
\includegraphics[width=0.4\textwidth]{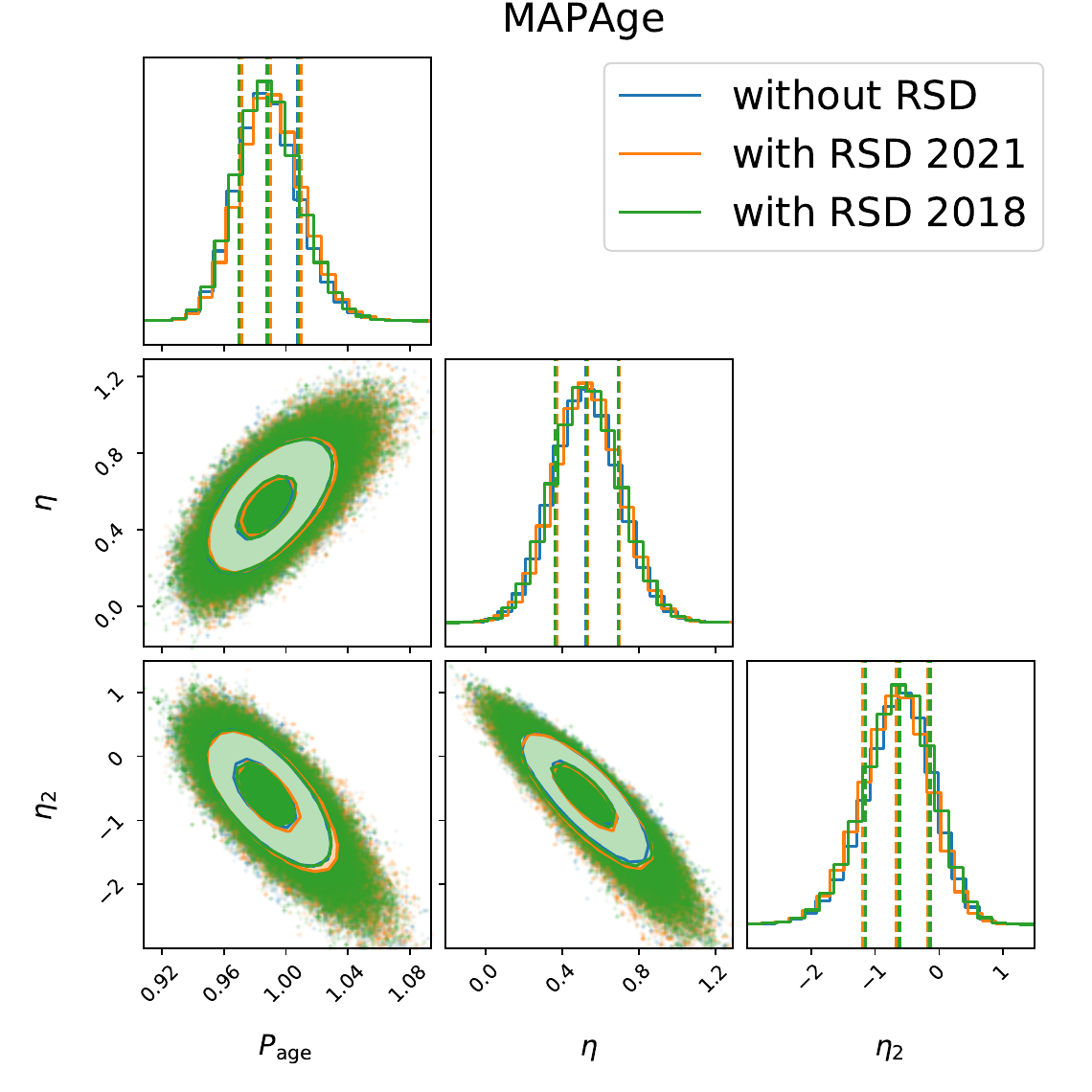}\\
\caption{Data comparison from fitting the background IDL without RSD data (blue) and the perturbative IDL data  (RSD+CC+BAO+SNe) with inclusions of RSD 2018 (green) and RSD 2021 (orange) data to the $\Lambda$CDM (top), PAge (medium), and MAPAge (bottom) models.}
\label{fig:datacomparison}
\end{figure}

It is easy to see that, with background IDL data, there is a positive evidence ($\Delta\mathrm{BIC}\simeq5$) against the late-time new physics parametrized by PAge model over the $\Lambda$CDM model, which is further strengthened when fitting to the perturbative IDL data involving with $S_8$ parameter that admits strong evidence ($\Delta\mathrm{BIC}>10$ for RSD 2018 and $\Delta\mathrm{BIC}>8$ for RSD 2021) against the PAge model over the $\Lambda$CDM model. Similar but slightly weaker conclusion could also be achieved for fitting the same model to the perturbative IDL data involving with $S_{12}$ parameter. In all cases, the MAPAge model is even more disfavored compared to the PAge model with $\Delta\mathrm{BIC}\simeq6$. Therefore, the final conclusion draw from the Tables~\ref{tab:background},~\ref{tab:S8}, and~\ref{tab:S12} is that, not only there is no need to complexify the late-time new-physics parametrization from PAge to MAPAge models, but also these late-time new physics parametrized by either PAge or MAPAge models are strongly disfavored compared to the $\Lambda$CDM model.

\subsection{Model comparison}

In Fig.~\ref{fig:modelcomparison}, models are compared via fitting the $\Lambda$CDM (blue), PAge (orange), and MAPAge (green) models to the background IDL data (CC+BAO+SNe) (top panel) and the perturbative IDL data (RSD+CC+BAO+SNe) with inclusions of the RSD 2018 (medium panel) and RSD 2021 (bottom panel) data. In the top panel, the late-time new-physics models parametrized by either PAge and MAPAge models are indistinguishable from the $\Lambda$CDM model, recovering our previous no-go guide for the Hubble tension in~\cite{Cai:2021weh}. In the medium and bottom panels, the late-time new physics from both PAge and MAPAge models is indistinguishable from the $\Lambda$CDM model in the $H_0-S_8$ plane, however, there is a significant shift in the central values along with widened uncertainties when $\Omega_\mathrm{m}$ is involved as seen from the $H_0-\Omega_\mathrm{m}$ and $S_8-\Omega_\mathrm{m}$ planes. 

This could be traced back to the fact that $\Omega_\mathrm{m}$ is absent in background evolution equations \eqref{eq:PAge} and \eqref{eq:MAPAge} for both PAge and MAPAge models, which is only sensitive to $\Omega_\mathrm{m}$ via matter perturbation growth equation \eqref{eq:Da} when the RSD data is included. Therefore, the constraints on $\Omega_\mathrm{m}$ for both PAge and MAPAge models come from the RSD data alone, which is certainly weaker than the constraints on $\Omega_\mathrm{m}$ for $\Lambda$CDM model from both background IDL data and RSD data simply because that the background IDL data contains much more data points than the RSD data. As for the shift in the central values of $\Omega_\mathrm{m}$ for PAge/MAPAge models with respect to the $\Lambda$CDM model, it is simply another reflection that the RSD data prefers lower $\Omega_\mathrm{m}$ than the background IDL data, similar to the case of the $S_8$ tension where the RSD data generally predicts lower $S_8$ value than the CMB data. This further suggests that the background IDL data plays a similar role as the CMB data.
However, this shift of late-time new-physics models from the $\Lambda$CDM model is strongly disfavored as indicated from their BIC values in Table~\ref{tab:S8}.

\subsection{Data comparison}

In Fig.~\ref{fig:datacomparison}, datasets are compared via fitting the background IDL without RSD data (blue) and the perturbative IDL data (RSD+CC+BAO+SNe) with inclusions of RSD 2018 (green) and RSD 2021 (orange) data to the $\Lambda$CDM (top panel), PAge (medium panel) and MAPAge (bottom panel) models. In the top panel, $S_8$ value is uplifted from $S_8=0.760^{+0.029}_{-0.028}$ to $S_8=0.811^{+0.027}_{-0.026}$ so that the usual $S_8$ tension disappears for the $\Lambda$CDM model when updating the RSD data from RSD 2018 dataset to RSD 2021 dataset as also shown in~\cite{Huang:2021tvo}. In the medium and bottom panels, the cosmological constraints on the PAge/MAPAge parameter space are stable regardless of datasets adopted for data fitting, therefore, both the background and perturbative IDL data are equally applicable in locating the PAge/MAPAge parameter space for the late-time new-physics models.

\section{Conclusions and discussions}\label{sec:con}

The Hubble tension is becoming a crisis for the recent $5\sigma$ claim from the recent local direct measurement, which carefully weights a comprehensive uncertainty budget for various systematics. If not caused by any other uncounted systematics, a full resolution for the Hubble tension would require for new physics beyond the $\Lambda$CDM model. Due to the large volume of proposed late-time solutions, it would be appealing to narrow down the possibilities in a model-independent manner. The traditional inverse distance ladder method is usually adopted to rule out the late-time homogeneous solutions, which could be further improved in a recent study~\cite{Cai:2021weh} by invoking a more accurate model-independent parametrization from the cosmic age and a cosmological-model-independent calibration to the traditional inverse distance ladder (BAO+SNe) from the cosmic chronometer. No appealing evidence beyond the $\Lambda$CDM model is found using this improved inverse distance ladder method, which is further strengthened in this paper when including the matter perturbation growth data. This suggests that if late-time solutions are invoked to address the new physics required for resolving the Hubble tension, then the modifications must go beyond the background level~\cite{DiValentino:2019ffd,Cai:2021wgv}. Our conclusion is consistent with recent studies ~\cite{Krishnan:2021dyb,Alestas:2021xes,Escamilla-Rivera:2021rbe,Ruiz-Zapatero:2022zpx}, together with the recent  criterion~\cite{Heisenberg:2022lob,Heisenberg:2022gqk,Lee:2022cyh}, rendering a no-go guide for the late-time solutions on the Hubble tension. 

For future perspective, recall that the PAge/MAPAge models neglect the short cosmic age spent during the radiation dominated era, therefore, it cannot parametrize the early-time solutions to the Hubble tension. However, it is feasible to use the exact solution describing the radiation-to-matter transition era,
\begin{align}
a(\tau)=a_\mathrm{eq}\left[\left(\frac{\tau}{\tau_*}\right)^2+2\left(\frac{\tau}{\tau_*}\right)\right],\, \tau_*=\frac{\tau_\mathrm{eq}}{\sqrt{2}-1},
\end{align}
where $\tau_\mathrm{eq}$ and $a_\mathrm{eq}$ are the conformal time and scale factor at matter-radiation equality, respectively. Then, a smooth transition from the radiation era with $Ht=1/2$ to the matter era with $Ht=2/3$ reads
\begin{align}\label{eq:Hteq}
Ht=\frac23\frac{\left(\tau/\tau_*+1\right)\left(\tau/\tau_*+3\right)}{\left(\tau/\tau_*+2\right)^2},
\end{align}
where the conformal time $\tau$ is solved from the integration $\int a(\tau)\mathrm{d}\tau$. Finally, the solution~\eqref{eq:Hteq} can be used to replace the factor of $2/3$ in the PAge model~\eqref{eq:PAge} to simultaneously characterize the early-time solutions from modifying the expansion history (changing $a_\mathrm{eq}$) and recombination history (changing $\tau_\mathrm{eq}$). We will investigate this new PAge model in future work to arrive at a full no-go guide for both early-time and late-time solutions to the Hubble tension.

\begin{acknowledgments}
This work is supported by the National Key Research and Development Program of China Grant No. 2021YFC2203004, No. 2021YFA0718304,
the National Natural Science Foundation of China Grants No. 12105344, No. 11647601, No. 11821505, No. 11851302, No. 12047503, No. 11991052, No. 12075297, No. 12047558,
the Key Research Program of the Chinese Academy of Sciences (CAS) Grant No. XDPB15, 
the Key Research Program of Frontier Sciences of CAS, 
the China Postdoctoral Science Foundation Grant No. 2021M693238, 
the Special Research Assistant Funding Project of CAS,
and the Science Research Grants from the China Manned Space Project with No. CMS-CSST-2021-B01.
We also acknowledge the use of the HPC Cluster of ITP-CAS.
\end{acknowledgments}


\bibliographystyle{utphys}
\bibliography{ref}

\end{document}